\documentclass[12pt]{article}
\usepackage{graphicx,color}
\usepackage[utf8]{inputenc}
\usepackage{a4,amssymb,amsmath}

\usepackage{newtxtext}

\numberwithin{table}{section}

\newtheorem{thm}{Theorem}[section]  
\newtheorem{prop}[thm]{Proposition} 
\newtheorem{lemma}[thm]{Lemma}       
\newtheorem{coro}[thm]{Corollary}   
\newtheorem{remk}[thm]{Remark}      
\newtheorem{note}[thm]{Notes}      
\newtheorem{defn}[thm]{Definition}  

\def\eref#1{(\ref{#1})}         
\def\sref#1{Sect.~\ref{#1}}
\def\aref#1{App.~\ref{#1}}
\def\pref#1{Prop.~\ref{#1}}
\def\lref#1{Lemma~\ref{#1}}
\def\rref#1{Remark~\ref{#1}}
\def\nref#1{Notes~\ref{#1}}

\def\tbref#1{Table~3.1}

\def\cref#1{Cor.~\ref{#1}}
\def\dref#1{Def.~\ref{#1}}
\def\fref#1{Footnote~\ref{#1}}

\def\bea#1{\begin{eqnarray}\label{#1}}
\def\eea{\end{eqnarray}}
\def\ba{\begin{array}}
\def\ea{\end{array}}
\def\qed{\hfill $\square$}

\def\RR{\mathbb{R}}

\def\HH{\mathcal {H}}

\def\OO{\mathcal{O}}
\def\WW{\mathcal{W}}
\def\VV{\mathcal{V}}
\def\SS{\mathcal{S}}
\def\gg{{\mathfrak{g}}}

\def\ka{\kappa}
\def\la{\lambda}

\def\pa{\partial}
\def\sumno{\sum\nolimits}
\def\inv{^{-1}}
\def\ol{\overline}

\def\ad{\mathrm{ad}}
\def\wt{\widetilde}

\def\sfrac#1#2{\hbox{\large{$\frac{#1}{#2}$}}}
\def\ndot{\!\cdot\!}
\def\Vert{\big\vert}
\def\nprod{\raisebox{1.5pt}{$\scriptstyle*$}}

\def\vev#1{\langle\!\langle #1\rangle\!\rangle}
\def\erw#1{\langle #1\rangle}
\def\Erw#1{\big\langle #1\big\rangle}
\def\wick#1{\colon\!\!#1\!\colon}

\numberwithin{equation}{section}

\title{Dressed fields for Quantum Chromodynamics}

\author{Iustus C. Hemprich$^{1}$
and Karl-Henning Rehren$^2$
\\[9pt]
{\small
$^1\,$Mathematisches Institut, Georg-August-Universität Göttingen,
37073 Göttingen, Germany} \\
{\small
$^2\,$Institut für Theoretische Physik, 
Georg-August-Universität Göttingen, 37077 Göttingen, Germany}}

\begin{document}

\maketitle

\begin{abstract}
String-localized QFT allows to explain Standard Model interactions in
an autonomous way, committed to quantum principles rather than a
``gauge principle'', thus avoiding an indefinite state space  and
compensating ghosts. The resulting perturbative scattering matrix is
known (at tree-level and without spacetime cutoff) to be insensitive
to the non-locality of the auxiliary ``string-localized free fields''
used in the construction. For the examples of Yang-Mills and QCD, we
prove that it is equivalent to the perturbative S-matrix of gauge
theory, restricted to physical  particle states. The role of classical
gauge invariance is revealed along the way. The main tool are
``dressed fields'', that are intermediate between free fields and
interacting fields, and for which we give explicit formulas at all
orders. The renormalization of loops, as well as non-perturbative
issues are not adressed, but we hint at the possibility, enabled by
our approach, that qualitative traces of confinement may be visible
already at the level of the dressed fields.

\end{abstract}

\section{Introduction}
\label{s:intro}
In a series of papers \cite{GMV,Gauss,Infra,AHM,GGM,GRV}, triggered by
the original ideas in \cite{MSY}, a group of researchers (including
one of the present authors) have developped and pursued an autonomous
approach \cite{aut} to the interactions of Standard Model (SM) particles.
``Autonomous'' means that it identifies admissible interactions by the imperative
that perturbative QFT must obey fundamental quantum principles:
locality of observables, Einstein causality, covariance, and the existence of a Hilbert
  space representation. While it is not by itself ``axiomatic'', the approach offers ways out
  of the notorious conflicts of gauge interactions with the
  strict axioms of axiomatic QFT, leading to physically most welcome
  broadenings of the latter.

Most prominently, one does not want to start from a gauge
principle to select the interactions, because quantization of gauge
field theories notoriously violates Hilbert space positivity. 
Sure enough, BRST is an established method in order to return to a 
Hilbert space, but it eliminates electrically charged interacting
fields: they would be (anti-)local in perturbation theory with a
strictly local interaction, but they lie outside the BRST
cohomology. On the other hand, in QED, by Gauss' Law charged fields 
must not commute with the electric field at spacelike infinity, so
they cannot be (anti-)local \cite{FMS}. This is a solid
physical reason to question the local gauge interaction plus BRST, and rather
relax the localization of charged fields.

The autonomous approach constructs charged fields that are localized
(in the sense of commutation relations with local observables) along ``strings''
extending to infinity. For this reason, the approach is also known as ``string-localized
QFT'' (sQFT).

In the case of QED \cite{Infra}, these fields do not commute
  with the electric field at infinity. They create
``infraparticle'' states, describing charged particles with attached
``photon clouds''. The clouds Lorentz transform nontrivially and carry
a rich superselection structure. These features of QED were anticipated
axiomatically \cite{Bu,FMS}. In particular, the Wightman
axioms of locality and covariance are too narrow to include infraparticle fields.

We briefly explain how sQFT achieves these features.
Recall that free local and covariant massless vector potentials
(for photons or gluons) do not exist on a Hilbert space, and free local
and covariant massive vector potentials (for $W$ and $Z$ bosons) exist
on a Hilbert space but do not admit renormalizable interactions. In
this situation, sQFT envisages (in a first step)
vector potentials $A_\mu(x,c)$ localized in spacelike
cones $x+\mathrm{supp}\,(c)$ (called ``string''), parametrized by a
suitable function $c^\mu(y)$, see \sref{s:Ac}. These fields are constructed in
both cases, massless and massive, on the Wigner Hilbert space of
helicity or spin 1, \textit{and}
they allow power-counting renormalizable interactions $L(c)$.  

Perturbation theory with string-localized vector potentials
  produces interacting fields on a Hilbert space whose properties can
  be understood in terms of ``dressed fields'', see \sref{s:dressed}.
  They are in general string-localized (in a less trivial way than the
  initial potentials $A_\mu(x,c)$ that are used only for setting up
  the interaction), and qualify as infraparticle fields, at least for
  QED. In more general theories, the construction is quite sensitive
  to possible obstructions \cite{GMV,AHM,GRV}. It is the main purpose
  of this paper to show how classical gauge invariance secures that it
  is consistently possible also in the case of QCD, without conflict
  with fundamental quantum physics. Of particular interest 
  will be the structure of the nonabelian dressing factor, \sref{s:main}.

The string is a necessity to evade pertinent mathematical NoGo theorems, 
but it should not have a meaning for the physical interaction. In
particular, the scattering matrix computed with $L(c)$ must not depend
on the string function $c^\mu$.
\footnote{Yet, matrix elements
    in infraparticle states will depend on the string functions of
    their clouds.}
This condition (``principle of string-independence'') turns out to restrict the 
form of the interaction as a polynomial in the Wick algebra of free
fields: there is a simple condition for the leading part in a coupling
constant $g$, and a hierarchy of further conditions, which require the
presence of higher-order (in $g$) interactions to \textit{cancel}
certain \textit{obstructions against string-independence} that arise
due to the time-ordering prescription in the defining expression for the S-matrix.

The surprise is that the interactions involving particles of
helicity or spin $1$, which are selected by these conditions, are
exactly those known in the SM \cite{GGM,GMV,GRV}, with the
string-localized potential supplanting the gauge potential.
Among other things, the massive $W$ and $Z$ bosons can be
described without a ``spontaneously broken gauge symmetry''. Instead,
string-independence requires the presence of and coupling to a Higgs
field fluctuating around one of the minima of the so-called Higgs
potential \cite{GRV}. It also requires chirality of weak interactions \cite{GMV}.

It is conceptually extremely comforting that the condition of
string-independence of the S-matrix, that is ultimately based in
quantum principles, has the same physical predictive power as the
gauge principle. Here, the obvious question suggests itself: Why is
this so? Perhaps, one should ask the other way round: Why is gauge
invariance so successful in QFT, despite its dubious ``negative
probabilities'' and ``ghosts''? 

The ``bottom-up'' method to answer this question is a
perturbative analysis to decide under which conditions two different
interaction densities $L$ and $K$ (say: string-localized and gauge)
give rise to the same S-matrix. It leads to an intricate recursive scheme: At
each order $g^n$ in the coupling constant, one must find a vector-valued
quantum field $U_n^\mu$, which together with properly adjusted 
interaction densities $L_n$ and $K_n$ of both approaches must cancel
(``resolve'') an obstruction, see \sref{s:ME}. We shall refer to the power series
$U^\mu(x)=\sum_n \frac{g^n}{n!} U^\mu_n(x)$ as the ``mediator field''.

There arise two challenges, a technical one, and one concerning the physics: the
former is to find an efficient recursive way to \textit{compute} the
obstruction arising at each order, and the latter is to check that it is
resolvable at each order, i.e., $L_n, K_n$ and $U^\mu_n$ exist. If it
is not resolvable, there is no equivalence,  
and the string-localized model is useless as a Hilbert space
description of the gauge interaction. Empirically, this does not happen
(precisely) for the interactions of the SM. One of the
purposes of the present work is to give a reason why this is so.

The former challenge was met in \cite{LV}, by presenting a closed
(all-orders) formula, called the ``LV equation'', that is equivalent 
to the equality of the S-matrices before the adiabatic limit (removal
of the spacetime cutoff) is taken, and from which the obstructions can
be computed  recursively at each order, in order to decide whether they are
resolvable, and determine the interactions and the mediator field at
the next order. 

In the present paper, we shall rather use a weaker form of the LV equation
pertaining to the adiabatic limit (no spacetime cutoff). We call it the
\textit{``Mediating Equation''}, because it secures the equality of
S-matrices of two different interactions, in the sense explained
in \nref{n:disc}(i) below. It is an implicit equation for the mediator
field $U^\mu$ in terms of $L$ and $K$. Of course, it has no solution in general,
unless the latter are appropriately in tune with each other. 

The Mediating Equation allows us to pass from the previous bottom-up
approach to a top-down approach: We can ``guess'' a closed (all-orders)
solution mediating between a string-localized interaction and a
local gauge interaction, and only have to verify its validity a
posteriori -- which is a comparatively easy task.
The idea is to read the Mediating Equation as a relation involving the
free fields of one description and ``dressed fields'' of the other description. 
Then, the classical gauge invariance of the local gauge Lagrangian
suggests the form of the dressed fields which solves the equation.
This observation reinstalls the \textit{classical} ``gauge
principle'' as a powerful prerequisite to ensure the string-independence of
the S-matrix of sQFT, without being implemented in the actual
\textit{quantum} field theory. 

The unveiling of classical gauge invariance as a sufficient condition
for consistency of our first-principles-based approach
is very much reminiscent
of an analogous result in \cite{PGI}, in which classical gauge invariance
was shown to be a sufficient condition for consistency at all orders
of a perturbative BRST structure, which in its turn secures a Hilbert
space on which the S-matrix can be defined. Major differences are
the absence of dressed fields and the fact
that interacting charged fields cannot be addressed in \cite{PGI}
because the BRST method is local, while -- as we show in this paper
-- interacting charged fields are necessarily string-localized. 

We shall demonstrate the scheme in the cases of Yang-Mills (\sref{s:YM}) and
QCD (\sref{s:QCD}), which are sufficiently demanding to convey the
nontriviality of the issue.

In \sref{s:LK}, we introduce the Mediating Equation between two
interactions, and the dressed fields as a tool to switch between both
descriptions in the perturbation theory of fields. We then present the
basics of perturbation theory in sQFT in \sref{s:basics}, and then show
how the Mediating Equation (with a minor but physically irrelevant
modification due to the appearance of null fields in gauge theory) can be
solved in YM and QCD. Some technical parts are deferred to appendices.

\begin{note}\label{n:disc}  
(i)
The ``S-matrix'' in the adiabatic limit is understood as a
formal power series in free fields and propagators, from which
matrix elements with sufficiently smooth many-particle states can be
extracted. It is not an operator, and ``unitary'' only in this formal sense.
As in other approaches, IR problems of matrix elements involving massless
particles still require special attention, not to speak about the convergence
of the perturbative expansion. Accordingly, all manipulations of the
S-matrix are understood in the sense of a formal series of formal
integrals. However, the physical feature that charged states are
infraparticle states with photon (or gluon) clouds is expected 
to bring about a similar cancellation of IR divergences  as
Faddeev-Kulish dressing factors, although different in the details \cite{Infra}.
Yet, a satisfactory scattering theory for infraparticles in terms of
asymptotic dressed fields is still lacking \cite{GRT,Infra}.
\\[1mm]
(ii)
``Fields'' are free quantum fields on $\RR^{1+3}$, or Wick polynomials
thereof (such as interaction densities $L$) or formal power series
(such as the vector-valued mediator field $U^\mu$ or the dressed fields).  
The free fields may be string-localized, constructed from local
  free fields as in \sref{s:basics}, defined on suitable Fock spaces.
  They are ``on-shell fields'' satisfying field equations, notably 
  the Klein-Gordon or wave equation. In particular, the gauge
  potentials $A_\mu$ are taken in the Feynman gauge and are defined on
  an indefinite state space. The free fields generate a Wick algebra
  $\WW$, isomorphic to  the classical algebra of field configurations, 
quotiented by the ideal of field equations.
\footnote{\label{f:quotient}The relevance of this quotient is apparent
  in the free Dirac Lagrangian $\ol\psi (i\gamma^\mu\pa_\mu-m)\psi$
  which is identically zero in the Wick algebra and in the classical
  quotient algebra; but it is of course not zero as a functional on field
  configurations, from which the equations of motion are to be
  derived.}
In \sref{s:Wick}, the Wick algebra $\WW$ will be extended to an
  algebra $\WW_s$ admitting further string integrations.
\\[1mm]
(iii) We establish the announced results {\em at tree-level} only, without
further mentioning. The validity of the same results at loop level may be
imposed as a ``renormalization condition'', to be secured by
exploiting the freedom of renormalization of loop-diagrams with
string-localized propagators at each order of perturbation
theory. This task is beyond the scope of the paper.
\\[1mm]
(iv)
We also emphasize that our method is genuinely perturbative. One would
thus expect that it is unable to address issues like confinement. But to
the contrary, we believe that first traces of confinement (like an
increase of the energy with the distance in states containing a quark
and an antiquark, see \rref{rk:IR}(iii)) can already be seen in terms of the
dressed fields, which are substantially easier to handle than the fully
interacting fields. While this hope is in fact a major motivation for the
present work, its actual implementation is beyond our scope.
\end{note}
We write (in most cases)
simply $XY$ for the (commutative, or anti-commutative 
in the case of fermionic fields) Wick product. 
The (anti-) commutative time-ordered product (T-product) is written as
$TX\ndot Y$, e.g., in \eref{omega}, understood as a linear combination
of Wick products with time-ordered two-point functions (propagators) as
coefficients, by Wick's theorem. The non-commutative operator product
(hardly ever appearing, e.g., in \eref{Bog}) is written as $X\nprod
Y$. Vacuum expectation values and propagators are displayed as
$\vev{X\nprod Y}$ and $\vev{TX\ndot Y}$.

\section{Two interactions producing the same S-matrix}
\label{s:LK}

The results of this section are not specifically restricted to sQFT.
But \cref{c:ME} is particularly useful in sQFT, e.g., in order to establish the
string-independence of an S-matrix constructed with a string-dependent
interaction density $L(c)$; or in order to establish that an S-matrix constructed
with a gauge interaction $K$ can as well be constructed directly on a
Hilbert space. The Mediating Equation \eref{ME} must admit a
solution, consisting of a pair of interaction densities $L$ and $K$ along with
a mediating vector-valued quantum field $U^\mu$. The existence of a
solution therefore characterizes the compatibility of different 
fundamental principles of QFT, which is usually not manifest. This 
condition strongly constrains admissible interactions. In particular, 
the leading (in the coupling constant) terms $L_1,K_1,U_1^\mu$
(which essentially specify a model) must
satisfy the simple ``initial condition'' \eref{LV}, and higher-order
interaction densities are determined by the solvability of \eref{ME}. 

\cref{c:dress} allows to control the localization of interacting
fields constructed with a string-localized interaction. For the
purposes of this work, it also paves the way to establishing the
necessary condition in \cref{c:ME} (as in \cite[App.~A.3]{LV}) and its
adaptation to gauge theory in \lref{l:MME}.

\subsection{Obstructions}
\label{s:obst}
We shall have to consider differences
  (``obstructions'') of the form
\bea{OYX} \notag
O_\mu(Y(x'),X(x)) &\equiv& [T,\pa'_\mu] Y(x')\ndot X(x)\big\vert^{\rm tree}
\\ &:=& \big( T\pa'_\mu Y(x')\ndot X(x)
- \pa'_\mu TY(x')\ndot X(x)\big)\big\vert^{\rm tree},\quad
\eea
and similar expressions with several factors $X_k(x_k)$. It follows from Wick's
theorem that 
\bea{OYXexp}
O_\mu(Y(x'),X(x)) =
\sumno_{\chi,\varphi} \wick{\sfrac{\pa Y}{\pa\chi}(x')\sfrac{\pa
  X}{\pa\varphi}(x)} \cdot O_\mu(\chi(x'),\varphi(x)), \eea
where the sum is over the linear generators of $\WW$ (counting
derivatives as independent fields), and the last factor is a numerical
difference of propagators. For local fields $\chi,\varphi$, it
is a distribution supported at $x'=x$, i.e., a linear combination of
derivatives of $\delta(x'-x)$.

Expressions like \eref{OYX} with several factors $X_k(x_k)$ can
  be expressed in terms of T-products involving $O_\mu(Y(x),X_k(x_k))$
  as factors. This is achieved by  the ``Master Ward Identity''  (MWI)
\begin{lemma}\label{l:MWI} {\rm (Master Ward Identity, \cite[Eq.~(43)]{BD}.)}
With the convention in \dref{d:outside}, it holds
\bea{MWI}
\big([T, \pa'_\mu] Y(x') \ndot X_1(x_1)\ndot \ldots \ndot X_n(x_n)\big)
\big\vert^{\rm tree}
= \sumno_k TX_1\ndot \ldots\ndot O_\mu(Y(x'),X_k(x_k))\ndot \ldots
\ndot X_n\big\vert^{\rm tree}.\,\,
\eea
\end{lemma}
\begin{defn}\label{d:outside} 
Whenever $O_\mu(Y(x'),X(x))$ appears as a factor in a T-product,
    the numerical distributional factors $O_\mu(\chi(x'),\varphi(x))$
    as in \eref{OYXexp} have to be taken outside the T-product.
\end{defn}
\dref{d:outside} takes care of a subtlety \cite{BD}: When \eref{MWI}
is smeared with test functions, the derivatives of $\delta$-functions
on the right-hand side are integrated by parts, hitting the test
functions and the fields. For the MWI to hold, one must take the derivatives
hitting the fields \textit{outside} the T-product.
\footnote{\label{fn:MWI}
In other words, integration by part is not allowed \textit{inside} the T-product. 
  E.g., when $O(Y(x'),X(x)) = Z(x')\ndot \pa'\delta$, the correct
  prescription to compute $\int dx' \, h(x') T\big(O(Y(x'),X(x))\big)
  \ndot W(x'')$ yields $-\pa h(x) TZ(x)\ndot W(x'') - h(x)\pa TZ(x)\ndot W(x'')$.
  With the derivative \textit{inside} the T-product, the second
  term would be $- h(x) T\pa Z(x)\ndot W(x'')$ instead. 
  The error is $\sim O(Z,W'')$, see \lref{l:omU}. \dref{d:outside} is
  equivalent to the formalization in \cite[Sect.~2]{BD} with the help
  of an ``external derivative'' $\wt\pa$ such that $\wt\pa h:=\pa h$ and
  $T\cdots\wt\pa Z\cdots := \pa T \cdots Z\cdots$.}
The convention \dref{d:outside} also applies to the following definition.
\begin{defn}\label{d:omega} {\rm(Obstruction map.)}
  For a hermitean vector-valued field $U^\mu$ of IR scaling dimension
\footnote{For massless fields, the IR dimension equals the UV
  dimension.
$\geq3$}, we define the obstruction map $\omega_U$ acting on the Wick
algebra $\WW$ as 
  \bea{omega}
\omega_U(X)(x)  := i\int dx'\, O_\mu(U^\mu(x'),X(x)).
\eea
\end{defn}
By the scaling dimension $\geq 3$, matrix elements of $U$ resp.\
  $\pa U$ in IR-regular states and contractions with interaction vertices
  fall off at large distance sufficiently fast such that the integral \eref{omega}
  exists, and $\int d^4x\,\pa_\mu U^\mu(x)=0$ in the subsequent arguments
  of \sref{s:ME}, i.e., integration by parts does not produce boundary terms.
  Moreover, $\omega_U$ can only increase the IR dimension of $X$.

In local QFT, with both $U^\mu$ and $X$ local fields,
  $\omega_U(X)(x)$ is localized at the point $x$, thanks to the
  $\delta$-functions.
  In sQFT, however, one has to extend the above to string-localized
fields. Basically, the numerical distributions $O_\mu(\chi(x'),\varphi(x))$ 
may involve string-integrated
$\delta$-functions, so that $\omega_U(X)(x)$ may be localized along the
string emanating from $x$. See \sref{s:PO} and \lref{l:omU}(i).

The map $\omega_U$ depends on the equations of motion and on the
propagators specified for the theory at hand (see \sref{s:PO}).

By \eref{OYXexp}, one easily sees that $\omega_U$ is a *-derivation 
w.r.t.\ the Wick product: $\omega(XY)=\omega(X)Y + X\omega(Y)$ and
$\omega(X^*)=\omega(X)^*$. It extends linearly to the Wick algebra of
power series of fields, when either $U^\mu$ or $X$ or both are power
series. As a consequence, its flow  $e^\omega$ is *-multiplicative:
$e^\omega(XY)=e^\omega(X)e^\omega(Y)$ and $e^\omega(X^*) =
e^\omega(X)^*$ (i.e., it is a *-morphism).

\subsection{The Mediating Equation}
\label{s:ME}

\begin{prop}\label{p:KLU}
  For Wick polynomials (or power series) $L(x)$, and $U^\mu(x)$ of
  IR dimension $\geq 3$, define
  \bea{KLU}
  K(x):= e^{\omega_U}(L)(x)-F(\omega_U)(\pa_\mu U^\mu)(x),
  \eea
where $F(\omega) = \frac{e^\omega-1}\omega = 1+
\frac12\omega+\frac16\omega^2+\dots$ as a power series of maps. 
Then the equality of S-matrices in the sense as specified in \nref{n:disc}(i)
follows:
\bea{S=S}
Te^{i\int dx\,L(x)}=Te^{i\int dx\,K(x)}.
\eea
\end{prop}
The proposition is a simplified version of \cite[Prop.~3.3]{LV},
pertaining to the adiabatic limit (i.e.~when the spacetime cutoff
function for the interactions is removed).
\footnote{The symbol $iO_U$ in \cite[Eq.~(3.14), (3.15)]{LV} is
  understood as
  $iO_{U(\chi)}$, which in the adiabatic limit $\chi\to 1$ becomes the
  present $\omega_U$. The condition in \cite[Prop.~3.3]{LV} secures 
  equality of S-matrices {\em before} the adiabatic limit is taken. It is
  equivalent to the present one in the case of simple models where all
  obstructions are multiples of $\delta$-functions (such as in
  \cite{AHM}). But it is manifestly stronger in models with non-abelian
  vector boson couplings, to the extent that it seems to be satisfyable only
  with sharp strings \cite[Sect.~3.6]{nab} -- which are expected to be
  too singular for the necessary IR regularization. Thus, in such models,
  the present weaker condition is more adequate. However, it
  secures equality of S-matrices only in the adiabatic limit.}

{\em Proof:} The interpolating family of interaction densities
$$
\ell_t(x) := e^{t\omega_U}(L)(x) - tF(t\omega_U)(\pa U)(x)
$$
satisfies the differential equation
$$
\pa_t\ell(x) = \omega_U(\ell_t)(x)-\pa U(x)
$$
with initial value $\ell_0(x)=L(x)$. Therefore,
$$
\frac{d}{dt} Te^{i\int dx\,\ell_t(x)} =i\int dx'\, T(\omega_U(\ell_t)(x')-\pa'
U(x'))\ndot e^{i\int dx\,\ell_t(x)}.
$$
Here, expanding the exponential, we use the Master Ward
Identity, \lref{l:MWI}. After subtraction of an integral over a 
derivative on the left-hand side, it implies 
\bea{expMWI}
i\int dx'\, T\pa' U(x') \ndot e^{i\int dx\,\ell_t(x)} = i\int
dx'\,T\omega_U(\ell_t)(x')\ndot e^{i\int dx\,\ell_t(x)}.
\eea
Thus,
$$
\sfrac{d}{dt} Te^{i\int dx\,\ell_t(x)} =0.
$$
With $\ell_0=L$ and $\ell_1=K$, the claim follows.
\qed

\pref{p:KLU} is usually of little worth, unless both $L$ and $K$ have
reasonable properties so as to qualify as interactions, such as
``Hilbert space'', ``locality'', or ``renormalizability''. An interesting
scenario is that these properties are complementary, e.g., $K$
is string-independent (hence local) but not defined 
on a Hilbert space, while $L$ is defined on a Hilbert subspace of the
space where $K$ is defined. Namely, in this case the S-matrix is
string-independent on the Hilbert space.  

The physically relevant reading of \pref{p:KLU} is thus not to
take \eref{KLU} as a definition of $K$ with given $L$ and $U^\mu$,
but rather as the challenge, for $L$ and $K$ subject to suitable
``reasonable properties'', to find $U^\mu$ mediating between them.
With this view, we reformulate the Proposition as a condition when two
interaction densities $L$ and $K$ give rise to the same S-matrix:
\begin{coro}\label{c:ME} {\rm(Mediating two interactions.)}
Two interaction densities $L$ and $K$ give rise to the same S-matrix,
i.e., \eref{S=S} holds in the sense as specified in \nref{n:disc}(i),
provided there exists a vector-valued ``mediator'' field $U^\mu$
solving the ``Mediating Equation''
\bea{ME}
e^{\omega_U}(L)-K = F(\omega_U)(\pa_\mu U^\mu).
\eea
\end{coro}
Because $\omega_U$ ``acts point-wise'' (or possibly ``string-wise''
in sQFT), the structure of the Mediating Equation is substantially
simpler than the direct comparison of the Wick expansions of two
S-matrices into Feynman integrals. In exchange, it involves the a
priori unknown mediating field $U^\mu$ and higher-order
  interaction densities $L_n$, $K_n$. It is the main result of this work, that
with some insight (in particular with $L$ and $K$ given independently
by \cite{GGM} and by gauge theory, respectively), this contrived task
can be accomplished by finding $U^\mu$.  

If $L$, $K$, $U^\mu$ are power series
\footnote{Typically, $L$ and $K$ are  polynomials, but $U^\mu$ is a
  non-terminating power series.}
in $g$ of leading order $\OO(g)$, then both \eref{S=S} and \eref{ME}
can be expanded as power series in $g$. The first-order condition for
both equations is simply 
\bea{LV}
L_1-K_1=\pa_\mu U_1^\mu,
\eea
i.e. the leading interaction densities must differ by a total derivative,
so that $\int dx\, L(x)=\int dx\, K(x)$. Higher orders of \eref{S=S} are more
involved, because in general $\int dx'\, T\pa'U'\ndot X \neq
\int dx'\,\pa' TU'\ndot X=0$, giving rise to ``obstructions'' $\omega_U(X)$.
The expansion of \eref{S=S} into time-ordered products can be recursively
brought into a form where at each order $L_n-K_n$ must cancel an obstruction
modulo a total derivative $\pa_\mu U_n^\mu$. E.g., the condition at
second order reads $L_2-K_2-\pa U_2= -\omega_{U_1}(L_1+K_1)$.
This is the old ``bottom-up'' approach. The Mediating Equation
\eref{ME} gives all higher-order conditions at one stroke. Thus, the
latter is in fact not only sufficient, but also necessary for the equality of S-matrices.  

The obstructions are resolvable when $L_n$, $K_n$ (subject to the respective
``reasonable properties'') and $U^\mu_n$ can be found to fulfill these
equations. Thus, the leading interaction densities $L_1$ and $K_1$, subject to
\eref{LV}, recursively determine $L$ and $K$ and $U^\mu$ to a large
extent.

Even when $L$ and $K$ have ``reasonable properties'', the interpolating
interaction densities $\ell_t$ (in the proof of \pref{p:KLU}) will in general
fail to have such properties. E.g., as \sref{s:main} will show, $L$
and $K$ may be renormalizable Wick polynomials, while $\ell_t$ are
non-terminating power series.

\subsection{Dressed fields}
\label{s:dressed}

Provided the Mediating Equation can be solved, a physically
  important by-product is the existence of ``dressed'' quantum fields.
In \sref{s:YM} and \ref{s:QCD}, we shall conversely use an
  ``educated guess'' for the dressed fields in the cases of YM and
  QCD, in order to prove that the Mediating Equation can be solved
  (with a slight modification necessary due to the presence of null
  fields, see \sref{s:MME}).

Dressed fields were first introduced in QED \cite{Infra}, where
  they account for the infraparticle nature of electrons and the
  associated superselection structure \cite{FMS}, mentioned in
  \sref{s:intro}. They are ``intermediate'' between the free and the
  interacting fields, see \eref{magic}. As will be explained below,
  they allow to assess in a straightforward way the localization of
  the actual interacting fields, which otherwise is at stake when the
  interaction is not strictly local. Yet another virtue of dressed fields will be discussed
  in \rref{rk:IR}(iii).

\begin{coro}\label{c:dress} {\rm(Dressed fields.)}
  Assume that \eref{ME}, hence \eref{S=S} holds. Define the respective
  interacting fields with interactions $L$ and $K$ by ``Bogoliubov's formula''
\bea{Bog}
  \Phi\big\vert_L(f) :=  -i\sfrac d{ds}\big\vert_{s=0} \big(Te^{i\int dx\,
  L(x)}\big)^*\nprod Te^{i\int dx\, (L(x) + sf(x)\Phi(x))},
\eea
  and likewise for $\Phi\vert_K(f)$. Then it holds
\bea{magic}
  \Phi\big\vert_L(x) = (e^{\omega_U}(\Phi))\big\vert_K(x).
\eea
  $e^{\omega_U}(\Phi)$ is called the ``dressed field'' associated with
  the free field $\Phi$.
\end{coro}
\textit{Proof.}
\eref{ME} continues to hold when $L$ and $K$ are replaced by
$L_s(x)=L(x)+sf(x)\Phi(x)$ and $K_s(x)=K(x)+sf(x)e^{\omega_U}(\Phi)(x)$.
Thus, one also has
$$
\big(Te^{i\int dx\, L(x)}\big)^*\nprod Te^{i\int dx\, L_s(x)} =
\big(Te^{i\int dx\, K(x)}\big)^*\nprod Te^{i\int dx\, K_s(x)}.
$$
The claim follows by taking the derivative $\frac d{ds}$ at $s=0$.
\qed

For examples of dressed fields in sQFT, see \cite{Infra,AHM,LV}, and
\sref{s:YM} and \ref{s:QCD}. The typical situation in sQFT is, that
the interaction density $L=L(c)$ and the mediator field $U^\mu=U^\mu(c)$
are string-localized, while $K$ is local. With the (even mild) non-locality
of $L(c)$, the usual Glaser-Epstein argument (based on the support of
retarded commutators) fails that warrants the relative locality of the
interacting fields among each other. Instead, one would have to study their
mutual commutation relations in terms of series in retarded integrals
involving interactions everywhere in the causal past. Here, the
dressed fields come to aid: one can see (manifestly, by the localization
properties of the map $\omega_U$, \sref{s:obst}) that the dressed
fields $e^\omega(\Phi)$
are string-localized relative to the free fields and to each other. Then,
because $K$ is local, the Glaser-Epstein argument suffices to conclude
that $e^\omega(\Phi)\vert_K$ are string-localized relative to each other.  
Thus, by \eref{magic}, also $\Phi\vert_{L(c)}$ are string-localized
relative to each other. This is the main virtue of \cref{c:dress}.

In particular, while local interacting charged fields do not exist on
the BRST Hilbert space of gauge theory, string-localized interacting
charged fields exist on the Hilbert space of sQFT.

\section{Basics of string-localized perturbation theory}
\label{s:basics}

\subsection{String-localized potentials}
\label{s:Ac}
Let $I_c^\mu$ stand for the convolution
\bea{Ic}
I_c^\mu(f)(x) := \int d^4y\, c^\mu(y) f(x+y) = \int d^4y\, c^\mu(y-x) f(y)
\eea
of a function or a quantum field of sufficiently rapid decay, with a
real vector-valued function or distribution $c^\mu$ on Minkowski
spacetime $\RR^{1,3}$, decaying towards infinity and satisfying
\bea{pac}
\pa_\mu c^\mu(y)=\delta(y).
\eea
Such $c^\mu$ necessarily must have a support extending to infinity,
and may be supported in some (arbitrarily narrow) cone, called a
``string'', extending from $0$ to infinity. For the purposes of sQFT,
the cone may be chosen spacelike.
\footnote{The present characterization \eref{pac} was already used by Dirac
  \cite{Dirac} and by Steinmann \cite{Stein}, for related but
  different purposes. In previous work of sQFT, more special forms of
  $c^\mu(y)$ have been employed. E.g., when $dc(e)$ is a smooth
  measure on a spacelike unit sphere $S^2\subset \RR^{1,3}$, then
  $c^\mu(y)  :=\int_{S^2} dc(e)\, e^\mu \int_0^\infty ds\, \delta(y-se)$
  satisfies \eref{pac} provided $\int_{S^2}dc(e)=1$, complies with
  Hilbert space positivity of the string-integrated field gauge vector
  potential \eref{phi}, see \rref{rk:IR}(i), and is sufficiently regular for a decent
  treatment of the infrared singularities of QED \cite{Infra}. It is supported
  in a narrow cone if $dc(e)$ is supported in a small solid angle.
  The present form is more flexible and may be useful to
    pass to curved background manifolds. 
  }
This operation (called ``string integration'') commutes with derivatives,
and it holds
\bea{Ipa}
I_c^\mu (\pa_\mu f)(x)= \pa_\mu (I_c^\mu)(x)= - f(x), \quad
\hbox{in short:}\quad (I_c\pa)=(\pa I_c)=-\mathrm{id}.
\eea
It is clear that if $X(x)$ is a local quantum field, then $I^\mu(X)(x)$
is localized in the string $x+\mathrm{supp}\,(c)$. 


Let $F_{\mu\nu}(x)$ be the local free quantum Maxwell tensor defined in
the Fock space $\HH$ over the unitary massless Wigner representations of
helicity $h=\pm1$. Then the hermitean string-localized field
\bea{Ac}
A_\mu(x,c):= I_c^\nu(F_{\mu\nu})(x)
\eea
is a potential for $F_{\mu\nu}$ defined on $\HH$:
$$
F_{\mu\nu}= \pa_\mu A_\nu(c)-\pa_\nu A_\mu(c),
$$
and is transversal and ``axial'':
$$
\pa_\mu A^\mu(c)=0, \qquad I_c^\mu (A_\mu(c))=0.
$$
These two relations hold by construction, rather than being imposed as
conditions ``in order to reduce the number of degrees of freedom'', as
in gauge theory. 

Perturbation theory is formulated in terms of an interaction density
$L(c)$, a Wick polynomial involving the potential $A_\mu(c)$. In the
case of QED, this is the minimal coupling $A_\mu(c)j^\mu$ of
the string-localized potential to the Dirac current \cite{Gauss,Infra}.

\subsection{String-localized Yang-Mills (sYM).}
\label{s:sYM}

String-localized Yang-Mills theory (sYM) has several copies of  $A_\mu(c)$,
one for each gluon, transforming in the adjoint representation of a
finite-dimensional Lie algebra $\gg$ of compact type. We introduce a basis
$\xi_a$ of $\gg$ with completely anti-symmetric real structure constants:
\footnote{We adopt the physicists' convention that in unitary
  representations $\pi(\xi_a)$ are hermitean matrices, so that the
  Lie-algebra bracket is $i[\cdot,\cdot]$.}
$$
i[\xi_a,\xi_b]=\sumno_cf_{abc}\xi_c,
$$
and the invariant quadratic form on $\gg$
$$
\erw{\xi_a\vert\xi_b}=\delta_{ab}, \qquad \erw{i[f,g]\Vert h} =
\erw{f\vert i[g,h]} \quad (f,g,h\in\gg).
$$
We conveniently combine quantum fields in the adjoint represention
into ``Lie-algebra valued'' fields
$$
X(x) = \sumno_a X_a(x)\xi_a, \quad \hbox{such that}\quad \sumno_a
X_aY_a= \erw{X\vert Y}.
$$

It was shown elsewhere \cite{Gass,GGM} that requiring the S-matrix
\bea{Sc}
S_{L(c)}=Te^{i\int dx\, L(x,c)}
\eea
to be independent of the string function $c^\mu$, determines the
interaction density to be of the form 
\bea{L}
L(c)=gL_1+ \sfrac{g^2}2L_2,&&
\\
\label{L12}
L_1= \sfrac12\Erw{F^{\mu\nu}\Vert i[A_\mu(c),A_\nu(c)]}, &&
L_2=-\sfrac12\Erw{i[A^\mu(c),A^\nu(c)]\Vert i[A_\mu(c),A_\nu(c)]}.
\eea
This result was obtained in the autonomous ``L-Q setting'', where the
invariance of $S_{L(c)}$ under variations $\delta_c$ of the string
function $c^\mu$ was addressed. It proceeds directly on the Wigner
Hilbert space where $F_{\mu\nu}$ and $A_\mu(c)$ are defined. One makes
an ansatz for a cubic renormalizable interaction density $L_1$, for
which string-independence 
of $S_{L(c)}$ at first order in $g$ implies the form given in \eref{L12}, and
the need to cancel an obstruction at second order requires the
``induced'' interaction density $L_2$ in \eref{L12}. No higher-order
interactions $L_n$ ($n>2$) are induced. $L(c)$ is identical with the
Yang-Mills interaction density of gauge theory, with $A$ replaced by $A(c)$.

Beyond the mere string-independence of $S_{L(c)}$, we establish its
matching with the (manifestly string-independent) S-matrix of the
gauge theory approach, by virtue of the Mediating Equation. 

We shall present exact formulas for the dressed fields in terms of a
group-valued field $W(x,c)=e^{i\gamma(x,c)}$. The mediator field will
be of the form
$$
U^\mu+ \wt U^\mu = \Erw{F^{\mu\nu}\Vert\alpha_\nu} +
\Erw{\beta\Vert j^\mu},
$$
where $\wt U^\mu$ is absent for Yang-Mills. The Lie-algebra-valued fields
$\alpha_\nu$, $\beta$ and $\gamma$ are determined by implicit equations,
which can be solved as power series expansions in $g$ (or in $A_\mu$).

\subsection{Gauge theory vs string-localized YM}
\label{s:YM-sYM}

We want to compare the S-matrix \eref{Sc} with the S-matrix of gauge theory,
$S_K=Te^{i\int dx\, K(x)}$ with the interaction density (without
ghosts and gauge-fixing)
\bea{K}
K=gK_1+ \sfrac{g^2}2K_2,&&
\\ \label{K12}
K_1= \sfrac12\Erw{F^{\mu\nu}\Vert i[A_\mu,A_\nu]}, &&
K_2=-\sfrac12\Erw{i[A^\mu,A^\nu]\Vert   i[A_\mu,A_\nu]}.
\eea
Here, $A$ is the Lie-algebra-valued gauge potential whose components
have the indefinite two-point function $\vev {A^a_\mu(x)\nprod
  A^b_\nu(x')} = -\delta_{ab}\eta_{\mu\nu}W_0(x-x')$.

For the sake of the comparison of S-matrices, we have to represent the
Hilbert space field $F_{\mu\nu}$ of \sref{s:Ac}, and along with it 
$A_\mu(c)$ given by \eref{Ac}, in the indefinite Fock space of the
gauge potential $A$. This is done by representing $F_{\mu\nu}$ as
$F_{\mu\nu}=\pa_\mu A_\nu-\pa_\nu A_\mu$, which preserves all
correlation functions. We also define the field
\bea{phi}
\phi(c):=I_c^\nu A_\nu.
\eea
It then follows from \eref{Ac} and \eref{Ipa} that
\bea{paphi}
A_\mu(c) -A_\mu = I_c^\nu(\pa_\mu A_\nu-\pa_\nu A_\mu) -A_\mu
= \pa_\mu\phi(c).
\eea
\begin{remk}\label{rk:IR}
(i)
It is crucial that the function $c^\mu(y)$ can be chosen such that
its values lie in a spacelike plane (see \cite{Infra} for the case of QED).
In this case, $\phi(c) =I_c^\mu (A_\mu)$ ``sees'' only three of the four
components of the indefinite gauge potential $A_\mu$, which generate a
positive-definite subspace of the gauge Fock space. The same will be true
for all fields in the space $\SS^\gg$, defined in \dref{d:SV}, which make up
the pertinent ``dressing factor'' $W(c)=e^{i\gamma(c)}$ in \sref{s:main}.
In particular, the dressed quark field \eref{psidress} constructed in
\sref{s:QCD} is defined on this subspace together with the Dirac Fock
space.
\\[1mm]
(ii)
The field $\phi(c)$ is infrared singular and needs to be regularized.
This was done in \cite{Infra} for QED, in a positivity-preserving way,
for its Wick exponential $e^{ig\phi(c)}$ which appears as a dressing
factor in the dressed Dirac field, while the linear field $\phi(c)$
itself ``disappears from the theory''. As we shall determine the
dressed quark field for sQCD in \sref{s:QCD}, the suggestion is to
regularize the corresponding dressing factor $W(c)$ rather than its
exponent $\gamma(c)$.
\\[1mm]
(iii)
We expect that, like the regularized dressed Dirac field in sQED captures
salient features of interacting QED (photon clouds and superselection
structure \cite{Infra}), the dressed quark field captures much of the
infrared features of QCD, possibly including confinement. Indeed, the
abelian dressing factors of electrons being Weyl operators, photon
clouds are coherent states, and the energy of the photon clouds of an 
electron-positron pair can be computed to contain the Coulomb
potential. In the nonabelian case, the exponent $\gamma(c)$
being non-linear in the free fields, the corresponding calculation
of the quark-antiquark potential will be much more demanding. Finding an
increase with the distance would be a signal of confinement already
at the dressed level. These steps will not be addressed in the present paper.
\end{remk}
Trying to apply \cref{c:ME}, one encounters the problem that $L_1-K_1$
fails to be a derivative, as required in \eref{LV} as the first-order
``initial'' condition.
The reason is that $F_{\mu\nu}$ fails to satisfy the free Gauss Law;
instead it holds
\bea{N}
\pa^\mu F_{\mu\nu} = -\pa_\nu N, \qquad \hbox{where}
\quad N:=\pa_\mu A^\mu .
\eea
$N$ is a ``null field'' with vanishing two-point function
$\vev{N(x)\nprod N(x')}=0$, and vanishing correlations with
$F_{\mu\nu}$ and $A_\nu(c)$. The same is true for the propagators. But
because it has non-vanishing correlations and propagators with $A_\mu$,
it cannot be ignored. 

We are thus forced to admit a weaker first-order condition than
\eref{LV}: one verifies by direct computation that
\bea{LVN}
L_1-K_1 = \pa_\mu U_1^{\mu} + \erw{\pa^\mu N\vert\alpha_1^\mu},
\eea
where
\bea{UN}
U_1^{\mu} = \Erw{F^{\mu\nu}\Vert \alpha_{1,\nu}} + \Erw{N\Vert
\alpha_1^\mu}, \qquad \alpha_{1,\nu} = \sfrac12
i[\phi,A_\nu(c)+A_\nu].
\eea
We shall adapt the Mediating Equation to this situation in \sref{s:MME}.
\paragraph{Notation.}
The notation $C\sim N$ (``$C$ is linear in the null field $N$'') will
mean that a field $C$ is a sum of  terms, each containing at least one
factor of the null field $N$ or $\pa N$. The latter needs not necessarily
be an explicit factor as in \eref{LVN} or \eref{UN}, but may appear in
the integrand of some (string) integration, such as $I_c(\erw{N\vert X})$.

\subsection{Subspaces and extensions of the Wick algebra}
\label{s:Wick}

We denote by $\WW$ the Wick *-algebra generated by the hermitean
fields $\phi_a$, $A_a$, $A_a(c)$, $F_a$ and $N_a$. For fields transforming in
the adjoint representation of $\gg$, we write $\sum_aX_a\xi_a\in\WW^\gg\equiv
\WW\otimes \gg$, see \sref{s:Ac}. $\WW^\gg$ is a Lie algebra by
$i[X,Y]=X_aY_b[\xi_a,\xi_b] = (\sumno_{ab}f_{abc}X_aY_b)\xi_c$.
(Recall that $X_aY_b$ is the commutative Wick product.) $\WW^0$ is the
subalgebra of fields in the trivial representation, such as
interaction densities. Clearly, $\erw{\WW^\gg\vert\WW^\gg}\subset \WW^0$.

\newpage

We denote by $\WW_s$ the smallest extension of the Wick algebra $\WW$,
which contains along with fields $Y(x)$ and $X(x)$ of total
  scaling dimension $d_X+d_Y\geq 1$ also the fields
$(I_cX)(x)Y(x)$ and $I_c(XY)(x)$. The latter have scaling dimension $\geq0$.

Where appropriate, we use the standard extensions $[g]$ and $[[g]]$
for algebras of polynomials and power series in the coupling constant $g$.

E.g., polynomial interaction densities $L\in gW^0[g]$ and the mediator field
$U^\mu\in g\WW^\gg_s[[g]]$ (see \sref{s:main}) are polynomials resp.\
power series without leading term $g^0$.

For later purposes, we define subspaces $\SS^\gg\subset\WW^\gg_s$ and
$\VV^\gg\subset\WW^\gg_s$  as follows. 
\begin{defn}\label{d:SV}
  (i) $\SS^\gg$ is the smallest subspace of $\WW^\gg_s$ that contains all
  elements of the form
  $$
  I_c^\nu\big(\ad_{is_k}\cdots \ad_{is_1}(A_\nu)\big)\quad (k\geq 0)
  \qquad\hbox{and}\qquad I_c^\nu\big(\ad_{is_k}\cdots
  \ad_{is_1}(\pa_\nu s_0)\big)\quad (k\geq 1)
  $$ when $s_0,\dots,s_k\in \SS^\gg$. 
  \\
  (ii)
  $\VV^\gg$ is the smallest subspace of vector fields in $\WW^\gg_s$ that
  contains $A_\nu$ and $\pa_\nu s$ ($s\in\SS^\gg$), closed under the
  action of $\SS^\gg$ by $\ad_{is}$. In other words, $\VV^\gg$ is spanned by
  elements of the form $$\ad_{is_k}\cdots 
  \ad_{is_1}(A_\nu)\quad (k\geq 0) \qquad\hbox{and}\qquad \ad_{is_k}\cdots
  \ad_{is_1}(\pa_\nu s_0)\quad (k\geq 0)$$  when $s_0,\dots,s_k\in \SS^\gg$.   
\end{defn}
In particular, $\SS^\gg$ and $\VV^\gg$ are ``generated'' by the field $A$ and
the operations $I_c$ (acting on vectors) and $\pa$ (acting on
scalars).

Because the polynomial degree is additive under the Lie commutator,
the recursive definition of $\SS^\gg$ is non-circular.
\footnote{One may define the subspaces of polynomial degree $n$
spanned by the indicated elements, where all $s_i$ belong to subspaces
of degree $\nu<n$. E.g., the subspaces of $\SS^\gg$ of degree $1$ and $2$
are spanned by $I_c(A)=\phi$, and $I(i[\phi,A])$ and
$I(i[\phi,A])+I(i[\phi,\pa\phi])=I_c(i[\phi,A])+I_c(i[\phi,A(c)-A])
=I_c(i[\phi,A(c)])$, respectively. The subspaces of $\VV^\gg$ of degree
$1$ and $2$ are spanned by $A$ and $A+\pa\phi=A(c)$, and
$i[\phi,A]$ and $i[\phi,A(c)]$, respectively.}
\begin{lemma}\label{l:SV}
  (i)
  All elements of $\SS^\gg$ are Lorentz scalars of scaling dimension $0$,
  and all elements of $\VV^\gg$ are Lorentz vectors of scaling dimension $1$.
  \\
  (ii)
  For $v_\nu\in  \VV^\gg$, $I_c^\nu v_\nu\in \SS^\gg$. For $s\in \SS^\gg$, $\pa_\nu s\in\VV^\gg$.
  In particular, $\SS^\gg = I_c(\VV^\gg)$.
  \\
  (iii)
  $\SS^\gg$ is a Lie subalgebra of $\WW^\gg$.
\end{lemma}
{\em Proof:} (i) and (ii) follow trivially from the definitions, and
by induction in $k$. Notice that due to \eref{pac} the operation $I_c$
lowers the dimension by 1.
\\
As for (iii), let $s_1,s_2\in\SS^\gg$. Then also
$$
i[s_1,s_2] = -I_c^\nu(\pa_\nu i[s_1,s_2]) = - I_c^\nu(i[\pa_\nu s_1,s_2]
+i[s_1,\pa_\nu s_2]) = I_c^\nu(i[s_2,\pa_\nu s_1]) -I_c^\nu(i[s_1,\pa_\nu s_2])
$$
is in $\SS^\gg$.
\qed

Having scaling dimension zero, elements of $\SS^\gg$ are as IR singular as
$\phi$, and the discussion in \rref{rk:IR}(ii) will apply to them as well.
Indeed, it is apparent from the dressing transformations presented in
\sref{s:YM} and \sref{s:QCD} that they are relevant only in exponential form.

\subsection{Propagators and two-point obstructions}
\label{s:PO}

By \eref{OYXexp}, in order to evaluate $O_\mu(Y',X)$ for $Y\in\WW$ and $X\in \WW$,
it suffices to know the ``two-point obstructions'' $O_\mu(\chi',\varphi)$ for the
linear generators, i.e., the differences of propagators: 
\bea{Ochph}
O_\mu(\chi',\varphi) =
\vev{T\pa'_\mu\chi(x')\ndot\varphi(x)}-\pa'_\mu\vev{T\chi(x')\ndot\varphi(x)}.
\eea 
The differences do not vanish in general, because $\pa'_\mu\chi(x')$
may be determined in terms of other fields by some equation of motion,
see the example \eref{OAF} below. We therefore need to know the
propagators, which in turn can be determined from the two-point functions
$\vev{\pa'_\mu\chi(x')\nprod\varphi(x)}=\pa'_\mu\vev{\chi(x')\nprod\varphi(x)}$.

The latter are obtained from
\footnote{Here and below, all two-point functions and propagators
  contain a factor $\delta_{ab}$ in the Lie algebra indices,
  which we do not display.}
$$
\vev{F_{\mu\nu}(x')\nprod F^{\ka\la}(x)}= \pa'_{[\mu}\delta_{\nu]}^{[\ka}\pa^{\la]}
W_0(x'-x), \qquad \vev{A_\mu(x')\nprod A^\ka(x)}=-\delta_\mu^\ka W_0(x'-x)
$$
(where $W_0$ is the canonical two-point function of a massless free scalar
field, and of course $\pa'=-\pa$) by the definitions
$$
A_\mu(c):= I^\nu_cF_{\mu\nu}, \qquad \phi(c):= I_c^\nu A_\nu, \qquad
N:=\pa^\mu A_\mu,
$$
which hold also for the two-point functions. E.g.,
$$
\vev{A_\mu(x',c)\nprod F^{\ka\la}(x)}=
I'^\nu_c(-\pa^{[\mu}\eta^{\nu][\ka}\pa^{\la]})W_0(x'-x) 
= (-\pa'_\mu I'^{[\ka}\pa^{\la]}-\delta_\mu^{[\ka}\pa^{\la]})W_0(x'-x).
$$
The propagators arise by replacing $W_0$ by the time-ordered two-point
function $T_0$ of the massless free scalar field. (We do not have to make use
of the option of further renormalization, mentioned after \dref{d:omega}.)
When the differences \eref{Ochph} are computed, the differential relations
$$
\pa^\mu\! F_{\mu\nu}=-\pa^\nu\!  N, \quad \pa_{[\mu}A_{\nu]}=\pa_{[\mu}A_{\nu]}(c)
=F_{\mu\nu}, \quad \pa^\mu\!  A_\mu=\pa^\mu\!  A_\mu(c)=N, \quad
\pa_\mu\phi(c)=A_\mu(c)-A_\mu
$$
(``equations of motion'') as well as the wave equation $\square X=0$ for
all linear fields,  have to be respected. This may result in $\delta$-functions
because  $\square T_0(z)=-i\delta(z)$ while $\square W_0(z)=0$. E.g.,
\bea{OAF} \notag
O_\mu(A'^\mu(c), F^{\ka\la}) = \vev{TN'\ndot F^{\ka\la}} - \pa'_\mu
\vev{TA'^\mu(c)\ndot F^{\ka\la}} \\ \notag = 0-\pa'_\mu (-\pa'^\mu
I'^{[\ka}\pa^{\la]}-\eta^{\mu[\ka}\pa^{\la]})T_0(x'-x)
\\  =\square' I'^{[\ka}\pa^{\la]}T_0(x'-x)=-I'^{[\ka}_c\pa^{\la]}\,i\delta(x'-x).
\eea
Proceeding in the same way, one arrives at \tbref{tb:YM} of two-point
obstructions, relevant for the YM model. Notice that $\phi$ and $N$ are
``inert'' in the sense that they do not contribute contractions as in
\eref{OYXexp} when they arise as factors in the field $Y$. In particular,
inertness of $N$ secures that if $Y\sim N$, then also $O_\mu(Y',X)\sim N$.
(The same is not true for $X$ because of row 5 of \tbref{tb:YM}.) 
\begin{table}\label{tb:YM}
$$
  \begin{tabular}{l||c|c|c|c|c|c|}
 & $F_b^{\ka\la}$ &$A_b^{\ka}(c)$ &$A_b^{\ka}$
 &$\phi_b$&$N_b$&$\pa^\ka N_b$  \cr\hline\hline
$\!\! iO_\mu(F'^{\mu\nu}_a,\cdot)$ & $\eta^{\nu[\ka}\pa'^{\la]}\delta^{ab}_{x'x}$
& $(\eta^{\nu\ka}-I^\nu_c\pa'^\ka)\delta^{ab}_{x'x}$ & 
$\eta^{\nu\ka}\delta^{ab}_{x'x}$ & $I^\nu_c\delta^{ab}_{x'x}$ & $0$&$0$ \cr
$\!\! iO_{[\mu}(A'_{a,\nu]}(c),\cdot)\!\!$  &$0$&$0$&$0$&$0$ &$0$&$0$\cr     
$\!\! iO_\mu(A'^\mu_a(c),\cdot)$ & $I_c'^{[\ka}\pa'^{\la]}\delta^{ab}_{x'x}$ 
    & $\!\big(I_c'^\ka -(I'_cI_c)\pa'^\ka\big)\delta^{ab}_{x'x}\!$
    &$I_c'^\ka \delta^{ab}_{x'x}$&$(I'_cI_c)\delta^{ab}_{x'x}$&$0$&$0$ \cr
 $\!\! iO_{[\mu}(A'_{a,\nu]},\cdot)$  &$0$&$0$&$0$&$0$ &$0$&$0$ \cr
 $\!\! iO_{\mu}(A'^\mu_a,\cdot)$ &$0$&$0$&$0$&$0$ &$-\delta^{ab}_{x'x}$
    & $-\pa^\ka\delta^{ab}_{x'x}$ \cr
 $\!\! iO_\mu(\phi'_a,\cdot)$ &$0$&$0$&$0$&$0$ &$0$&$0$\cr  
 $\!\! iO_\mu(N'_a,\cdot)$ &$0$&$0$&$0$&$0$ &$0$&$0$\cr  \hline
  \end{tabular}
  $$
\caption[Table 3.1]{Two-point obstructions in the gluon sector.
    ($\delta^{ab}_{x'x}\equiv \delta_{ab}\delta(x'-x)$) }
\end{table}
In nonabelian models of sQFT, there arise subtle issues with $O_\mu(Y',X)$
when $Y$ or $X$ belong to $\WW_s$ but not to $\WW$. In particular,
the expansion \eref{OYXexp} in the first entry  is not applicable, when $Y$ 
is of the form $Y=I_c(Y_1Y_2)$. In \aref{a:YM}, we shall discuss and tackle
these issues for the classes of string-integrated Wick products that
are pertinent for our main result.

\section{Matching of S-matrices for Yang-Mills}
\label{s:YM}
Due to the violation \eref{LVN} of the necessary ``initial condition''
\eref{LV}, we have to adapt \pref{p:KLU} and \cref{c:ME}. Namely, we
present a ``modified Mediating Equation'' \eref{MME} securing the
validity of 
\bea{SNS}
T e^{i\int dx\, (L(x,c) + C[N](x,c))} = T e^{i\int dx \, K(x)}
\quad\hbox{with}\quad C[N]\sim N. 
\eea
Admitting the modification \eref{SNS} is physically reasonable,
because the $N$-dependent term $C[N]$ on the left-hand side is
ineffective when the S-matrix is evaluated in states of the gluon
fields of the form $p_{[\mu}a_{\nu]}^*(p)\vert 0\rangle$, generated by
$F_{\mu\nu}$ acting on the Fock vacuum. This subspace is positive
semi-definite, and contains the two Wigner states of massless free
gluons, along with the ``longitudinal'' null states $p^\mu a_\mu^*(p)
\vert0\rangle$. The reason is that all contractions of $N$ with
$F_{\mu\nu}$ vanish, see \sref{s:YM-sYM}, hence all contractions with
the states and with the interaction density $L$ vanish. The only contractions
schemes that would contract all $N$ fields in the S-matrix either
involve the self-propagator $\vev{TN\ndot N}=0$, or produce a loop of
contractions of $N(x_i)$ with other fields within $(C[N])(x_j)$. Because
our analysis is restricted to tree-level, these schemes can be ignored. Thus,
on the subspace of Wigner states, \eref{SNS} is equivalent to \eref{S=S}.
In particular, the precise form of $C[N]\sim N$ is not relevant.

In the sequel, we refrain from writing the argument $c$, as in $L\equiv L(c)$
or $\phi\equiv \phi(c)$ -- except in $A_\mu(c)$ (the string-localized potential)
and $A_\mu$ (the gauge potential). We restore it in propositions.

\subsection{The Modified Mediating Equation}
\label{s:MME}

\begin{lemma}\label{l:MME} {\rm (Modified Mediating Equation.)}
  Assume that, for given $L$ and $K$, there exist fields $U^\mu$ of IR
  dimension $\geq 3$ and
$D[N]\sim N$ in $\WW_s$ solving the Modified Mediating Equation 
\bea{MME}
e^{\omega_U}(L) - K = F(\omega_U)(\pa U) - D[N].
\eea
If $\omega_U(N)\sim N$ and $\omega_U(\pa N)\sim N$, then \eref{SNS}
holds (again in the sense as specified in \nref{n:disc}(i))
with $C[N]:= e^{-\omega_U}(D[N])\sim N$. 
\end{lemma}
\textit{Proof.} By the derivation property of $\omega_U$, it follows that
$\omega_U(D[N])$ is again linear in $N$, and by iteration, $C[N]= 
e^{-\omega_U}(D[N])\sim N$. Then, \eref{MME} implies \eref{SNS} by \cref{c:ME}
with $L+C[N]$ supplanted for $L$ in \eref{ME} and \eref{S=S}. 
\qed

We shall actually solve \eref{MME} for sYM (i.e., $L=L(c)$ and $K$ as specified
in \sref{s:basics}) with $D[N]$ of the form $\erw{\pa^\nu  N\vert d_\nu}$
(proof of \cref{c:YM}) and $\omega_U(N)=\omega_U(\pa N)=0$ (\lref{l:omU}(i)),
hence $C[N]=\erw{\pa^\nu N\vert e^{-\omega_U}(d_\nu)}$.
The precise form of $D[N]\sim N$ is not relevant, since
$C[N]\sim N$ is not relevant  in \eref{SNS}. 

 We shall solve the modified Mediating Equation \eref{MME}, by
 ``guessing'' an expression for $e^{\omega_U}(L)$, from which $U^\mu$
 can be reconstructed. The guess is suggested by classical
 gauge invariance of the \textit{full} Yang-Mills Lagrangian $K_0+K$,
 including the quadratic part
 $$
 K_0:= -\sfrac14 \Erw{F^{\mu \nu} \Vert F_{\mu \nu}}.
 $$
Recall that the S-matrix involves only the interaction part $K$
displayed in \eref{K}. Let
$$
G_{\mu \nu} := F_{\mu \nu} - i g [A_\mu,A_\nu], \qquad G_{\mu \nu}(c)
:= F_{\mu \nu} - i g [A_\mu(c),A_\nu(c)].
$$
Then
$$
K_0+K =  -\sfrac14 \Erw{G^{\mu \nu}\Vert G_{\mu \nu}},\quad K_0+L =  -\sfrac14
\Erw{G^{\mu \nu}(c) \Vert G_{\mu \nu}(c)}.
$$
Using $\omega F(\omega) = e^\omega-1$, we rewrite \eref{MME} as
\bea{MME1}
e^{\omega_U}(L+K_0) = (K+K_0) + F(\omega_U)(\pa U+\omega_U(K_0)) - D[N].
\eea
The announced ``guess'' suggested by classical gauge invariance of
$K+K_0$ is that $e^\omega(G_{\mu\nu}(c))$ is an operator-valued gauge
transform of $G_{\mu\nu}$, so that
\bea{K0LK}
e^{\omega}(K_0+L)=K_0+K
\eea
will hold. Then, \eref{MME1} follows, provided $D[N]:=F(\omega_U)(\pa U
+\omega_U(K_0))$ is $\sim N$. This will be established in \pref{p:FAcdress}.

\subsection{The main result}
\label{s:main}

\pref{p:FAcdress} and \cref{c:YM} constitute the main result of this
paper, together with their analogs \pref{p:psidress} and \cref{c:QCD} for QCD.

Let the subspaces $\SS^\gg$ and $\VV^\gg$ of $\WW^\gg_s\subset\WW_s$ be defined as
in \sref{s:Wick}.
\begin{prop}\label{p:FAcdress} {\rm (YM dressing.)}
  (i)
  There exists a mediator field $U^\mu(c)\in  g\VV^\gg[[g]]$ of the form
\bea{Ualpha}
U^\mu(x,c)=\Erw{F^{\mu\nu}(x)\Vert \alpha_\nu(x,c)}
\eea
with $\alpha_\nu(c)\in g\VV^\gg[[g]]$, which generates the dressing transformations
\bea{Acdress}
e^{\omega_U}(A_\mu(x,c)) &=& W(x,c)\big(A_\mu(x)+ i g\inv \pa_\mu\big)W(x,c)\inv,
\\ \label{Fdress} 
e^{\omega_U}(F_{\mu\nu}(x))
&=& \pa_\mu e^{\omega_U}(A_\mu(x,c))
-\pa_\nu e^{\omega_U}(A_\mu(x,c)),
\eea
where
\bea{Wgamma}
W(x,c)=e^{i\gamma(x,c)} \quad\hbox{with}\quad \gamma(c)\in g\SS^\gg[[g]]
\eea
is a group-valued string-localized field. $W\inv =e^{-\gamma}$ is
the inverse w.r.t.\ the Wick product.
\newpage

(ii)
It holds $\omega_U(N)=\omega_U(\pa N)=0$, and 
\bea{UomK}
\pa_\mu U^\mu(x,c)+\omega_U(K_0(x)) = -\Erw{\pa^\nu N(x)\Vert \alpha_\nu(x,c)}.
\eea
\end{prop}

In order to prove \pref{p:FAcdress}, the task is to find $\gamma(c)$
along with $U^\mu(c)$ with the stated properties. This is postponed after
the proof of \pref{p:Acdress}.

The proof of the following corollary to \pref{p:FAcdress}
reveals the role of classical gauge invariance.
\begin{coro} \label{c:YM} {\rm (Equality of S-matrices in YM.)}
 The Modified Mediating Equation  \eref{MME} is satisfied, and the
 equality \eref{SNS} of S-matrices holds for the case of Yang-Mills theory. 
  \end{coro}
  \textit{Proof.}
Assume that \pref{p:FAcdress} holds. Thinking of $K_0+K = F[A]$ as a
functional that depends on $A_\mu$ only through $G_{\mu\nu}$, we have
$K_0+L=F[A(c)]$. \pref{p:FAcdress}(i) implies 
$$
e^{\omega_U}(K_0+L) = e^{\omega_U}(F[A(c)]) =  F[e^{\omega_U}(A(c))] =
F[W(A+ig\inv\pa)W\inv] = F[A]=K_0+K,
$$
where \eref{Fdress} and the multiplicative property of the map
$e^{\omega_U}$ secure the second equality and the third equality is
\eref{Acdress}. The fourth equality is the gauge invariance of
  the classical Lagrangian $K+K_0$, that holds as well when the classical
field $A$ and the gauge function $\theta$ in \eref{GT} are replaced by the quantum
field $A$ and the operator-valued function $W\inv$ under the Wick
product. This establishes \eref{K0LK}.
On the other hand, applying $F(\omega_U)$ to \eref{UomK} yields
$$
F(\omega_U)(\pa_\mu U^\mu)+e^{\omega_U}(K_0)-K_0 = -\Erw{\pa^\nu
  N\Vert F(\omega_U)(\alpha_\nu)},
$$
where \lref{l:omU}(ii) was used. Together with \eref{K0LK}, the
modified Mediating Equation \eref{MME} is obtained with
$D[N] = -\erw{\pa^\nu N\vert F(\omega_U)(\alpha_\nu)}$.
Thus, \lref{l:MME} applies.
\qed

To prepare the proof of \pref{p:FAcdress}, we claim
\begin{prop}\label{p:Acdress} {\rm (Determination of the dressing
    factor $W(c)$ and the mediator field of YM.)} 
\\  (i)
  The ansatz \eref{Acdress} implies
\bea{IAcdress}
I_c^\mu\big(W(c)\big(gA_\mu+ i \pa_\mu\big)W(c)\inv\big)=0.
\eea
(ii)
\eref{IAcdress} determines a unique power series $\gamma(c)\in g\SS^\gg[[g]]$
such that $W(x,c)=e^{i\gamma(x,c)}$.
\\ (iii)
Given $\gamma(c)$ from (ii),  \eref{Acdress} determines a
(non-unique) power series $\alpha(c)\in g\VV^\gg[[g]]$, such that
$\omega_U$ arises from $U^\mu(c)$ as given in \eref{Ualpha}.
\end{prop}

\textit{Proof.} The ``reconstruction'' of $\gamma$ and $\alpha$ from
\eref{Acdress} uses several results proven in \lref{l:omU}, where we assume
(and verify a posteriori) that $\gamma\in g\SS^\gg[[g]]$ and $\alpha\in g\VV^\gg[[g]]$.

(i)
We use \lref{l:omU}(iii) to conclude from
$I_c^\mu(A_\mu(c))=0$ that $I_c^\mu$ applied to the left-hand side of
\eref{Acdress} vanishes. This yields \eref{IAcdress}.
\\ (ii)
We write the right-hand side of the dressing transformation \eref{Acdress} as
\footnote{\label{fn:duh} using a standard result for derivations acting on
exponentials: $\delta(\exp X) = F(\ad_X)(\delta(X))\exp X$ with $F$
the same function as in \pref{p:KLU}, originally due to Duhamel (1856)
and Schur (1891 for Lie algebras)}
\bea{Acdress-r}
e^{i\gamma}(gA+ i \pa_\mu)e^{-i\gamma}= \exp(\ad_{i\gamma})(gA_\mu)
+ F(\ad_{i\gamma})(\pa_\mu\gamma) = gA_\mu + \pa_\mu\gamma
+ g\Sigma_\mu(\gamma),
\eea
where we have isolated the leading terms $(\ad_{i\gamma})^0$, and set
$$
\Sigma_\mu(\gamma):= \sumno_{n\geq 1}
\sfrac1{n!}\ad_{i\gamma}^n\big(A_\mu + \sfrac{g\inv}{n+1}\pa_\mu
\gamma\big)=\OO(g).
$$
We apply $I_c^\mu$ and use \eref{Ipa}: $I_c^\mu\pa_\mu\gamma
=-\gamma$. This yields the implicit equation for $\gamma$
\bea{gammaAZ}
\gamma =  I_c^\mu \big(gA_\mu+g\Sigma_\mu(\gamma)\big)
= g\phi + I_c^\mu\big(g\Sigma_\mu(\gamma)\big). 
\eea
When the expansion $\gamma=\sum_{n\geq1}\frac{g^n}{n!}\gamma_n$ is
inserted on both sides, the term of order $g^n$ on the right-hand side
contains only $\gamma_k$ with $k<n$. This yields recursive formulas
for $\gamma_n$ in terms of lower coefficients. At lowest orders:
\bea{gamma123} \notag 
& \gamma_1=I_c^\nu(A_\nu)\equiv \phi, \quad \gamma_2 =
I_c^\nu(\ad_{i\gamma_1}(2A_\nu+\pa_\nu\gamma_1)
\equiv I_c^\nu(i[\phi, A_\nu(c)+A_\nu]),
\\ 
& \gamma_3=I_c^\nu\big(\sfrac32\ad_{i\gamma_1}(\pa_\nu\gamma_2)
+ \ad_{i\gamma_2}(3A_\nu+\sfrac32\pa_\nu\gamma_1)
+ \ad_{i\gamma_1}\ad_{i\gamma_1}(3A_\nu+\pa_\nu\gamma_1) \big).
\eea
$\gamma_3$ can be further worked out by inserting $\gamma_1$ and
$\gamma_2$, cf.\ \cite[Eq.~(3.90)]{nab}.

One can prove by induction in $n$, that each $\gamma_n\in\SS^\gg$ (see
\dref{d:SV}), and that each $\gamma_n$ is homogeneous of degree $n$ in
$A$, with actions of derivatives and $I_c$.
\\
(iii)
Returning to \eref{Acdress}, we write $e^{\omega_U}(A(c))$ on
the left-hand side as $A(c) + F(\omega_U)(\omega_U(A(c)))$, and 
use \lref{l:omU}(i) to evaluate $\omega_U(A(c))$:
\bea{Acdress-l}
e^{\omega_U} (A_\mu(c))= A_\mu(c)+F(\omega_U)((\delta_\mu^\nu
+ \pa_\mu I_c^\nu)(\alpha_\nu)).
\eea
On the right-hand side of \eref{Acdress}, we use \eref{Acdress-r} and
\eref{gammaAZ}. Thus, \eref{Acdress} becomes
\bea{FpalphaZ}
F(\omega_U)\big((\delta_\mu^\nu + \pa_\mu I_c^\nu)(\alpha_\nu)\big) = 
(\delta_\mu^\nu + \pa_\mu I_c^\nu)(\Sigma_\nu(\gamma)),
\eea
where we have cancelled $A_\mu(c)$ against $(\delta_\mu^\nu + \pa_\mu
I_c^\nu)(A_\nu) = A_\mu + \pa\phi$. This is an implicit
equation for $\alpha$, because $\omega_U$ on the left-hand side
depends on $\alpha$, while $\Sigma_\mu(\gamma)$ is independent of
$\alpha$. Due to \eref{Ipa} and \lref{l:omU}(i), the left-hand side is invariant
under $\alpha_\mu\to\alpha_\mu+\pa_\mu s$ ($s\in g\SS^\gg[[g]]$), so the
solution is not unique. We fix the freedom by solving instead the
implicit equation
\bea{FalphaZ}
F(\omega_U)(\alpha_\mu) = \Sigma_\mu(\gamma),
\eea
which implies \eref{FpalphaZ} by virtue of \lref{l:omU}(ii) and
(iii). We rewrite \eref{FalphaZ} as 
$$
\alpha_\mu = \Sigma_\mu(\gamma) + (1-F(\omega_U))(\alpha_\mu) =  
\sumno_{n\geq1} \sfrac1{n!}\big(\ad_{i\gamma}^n
(A_\mu + \sfrac{g\inv}{n+1}\pa_\mu \gamma)
-\sfrac1{n+1}\omega_U^n(\alpha_\mu)\big).
$$
When the expansion $\alpha=\sum_{n\geq1}\frac{g^n}{n!}\alpha_n$ is
inserted, the term of order $g^n$ on the right-hand side contains only
$\alpha_k$ with $k<n$ (because $\omega_U$ is linear in $\alpha$).
This yields recursive formulas for $\alpha_n$ in terms of lower coefficients
along with the coefficients of $\gamma$, known from (ii). At lowest orders:
\bea{alpha12} \notag
& \alpha_{1,\mu} = \ad_{i\gamma_1}(A_\mu+\sfrac12\pa_\mu\gamma_1)
\equiv \sfrac 12 i[\phi,A_\mu(c)+A_\mu],
\\ 
& \alpha_{2,\mu}=\ad_{i\gamma_1}\ad_{i\gamma_1}(A_\mu
+\sfrac13\pa_\mu\gamma_1) + \ad_{i\gamma_2}(A_\mu+\sfrac12\pa_\mu\gamma_1) +
\sfrac12\ad_{i\gamma_1}(\pa_\mu\gamma_2) - \omega_{U_1}(\alpha_{1,\mu}),
\quad \eea
where $\omega_{U_n}$ is the contribution from
$U^\mu_n=\erw{F^{\mu\nu}\vert \alpha_{n,\nu}}$, according to \eref{Ualpha}.
This can be further worked out by inserting $\alpha_1$, $\gamma_1$, and
$\gamma_2$, and using \lref{l:omU} and \tbref{tb:YM} for the last term.

One can prove by induction that all $\alpha_n\in \VV^\gg$, hence $U^\mu$
has dimension 3 by \lref{l:SV}(i).

With the expansion of \eref{Acdress} satisfied at all orders in $g$,
\pref{p:Acdress} is proven.
\qed
\begin{remk}\label{rk:WA} $W(c)$ being the unique solution to \eref{IAcdress}
  of the form \eref{Wgamma}, may be
  regarded as a functional $W[A,I_c]$. Since each $\gamma_n$ is
  a Wick polynomial of degree $n$ in $A$ (with operations
  $I_c$), the power series expansions of $\gamma(c)\in\SS^\gg[[g]]$ 
  and $W(c)$ are actually expansions in the field $gA$.
\end{remk}
The proof of \pref{p:FAcdress} is now easily completed.

\textit{Proof of \pref{p:FAcdress}.}
(i) 
\eref{Ualpha} and \eref{Acdress} were already established in
\pref{p:Acdress}. \eref{Fdress} follows by \lref{l:omU}(ii) and
$F_{\mu\nu}=\pa_{[\mu}A_{\nu]}(c)$.
\\
(ii)
The first statement is part of \lref{l:omU}(i). Also by
\lref{l:omU}(i), we have
$$
\omega_U(K_0) = -\sfrac12 \Erw {F^{\mu\nu}\Vert\omega_U(F_{\mu\nu})}
= -\Erw{F^{\mu\nu}\Vert \pa_\mu \alpha_\nu}.
$$
Then \eref{UomK} follows from $\pa_\mu U^\mu = \erw{F^{\mu\nu}\vert
  \pa_\mu \alpha_\nu} - \erw{\pa^\nu N\vert \alpha_\nu}$.
\qed

In \aref{a:DFM}, we make contact with the ``Dressing Field Method'' \cite{DFM}
developped for classical gauge field theory. In a sense qualified there,
our quantum dressing transformation is closely related to the 
``dressing''  used in that approach. A major link is the following observation.  

Classical gauge transformations \eref{GT} can be lifted to the quantum
  field spaces $\SS^\gg$ and $\VV^\gg$, whose elements are viewed as
  functionals of $A$ and $I_c$,  by acting on the quantum field $A$
  with the same formula \eref{GT}. In particular, they act on $W(c)\in
  \exp ig\SS^\gg[[g]]$. There arises no conflict with quotienting by the wave
  equation, because the wave operator never comes to bear in \dref{d:SV}. 
\begin{prop}\label{p:GT-W} {\rm (Gauge invariance of the dressed
    field.)}
  Under the lifted gauge transformations \eref{GT}, the dressing factor
$W(c) \equiv W[A,I_c] \in \exp ig\SS^\gg[[g]]$ transforms as
  \bea{GT-W}
  W[A,I_c](x)\mapsto W[A^\theta,I_c](x) = W[A,I_c](x)\theta(x).
  \eea
Consequently, the dressed field $e^{\omega}(A_\mu(c))$ as given
  in \pref{p:FAcdress} is gauge invariant.
\end{prop}
\textit{Proof.}
By \rref{rk:WA}, $W[A,I_c]$ is the unique solution to the implicit
equation \eref{IAcdress}, that we write as
$I_c\big(A^{W\inv}\big)=0$, hence $W[A^\theta,I_c]$ is the unique 
solution to  $I_c\big((A^\theta)^{W\inv}\big)=0$. Because
$(A^\theta)^{W\inv}=A^{\theta W\inv}$, the same equation is solved
by $W[A,I_c]\theta$. This proves \eref{GT-W}. Gauge invariance of
$e^\omega(A(c))=ig\inv A^{W\inv}$ follows from
$(A^\theta)^{(W\theta)\inv}= A^{W\inv}$. 
\qed

\subsection{``Smeared Wilson operators''}
\label{s:Wilson}

The transformation law \eref{GT-W} is that of a Wilson operator
(parallel transporter) along a curve extending from $x$ to infinity
(under gauge transformations that are trivial at infinity). Indeed,
one can see (for the lowest orders of $\gamma(c)$ displayed above)
that when the function $c^\mu$ in \eref{Ic} is supported along a curve
$C_0$ from $0$ to infinity, then $W(c)$ is the proper Wilson operator
(path-ordered exponential) $Pe^{i\int_{C_x} A_\mu(y) dy^\mu}$ along
the curve $C_x=C_0+x$: The nested string integrations account for the
necessary permutations in the path-ordering. Since we allow
\footnote{as we must in QFT, because fields are distributions}
higher-than-one-dimensional supports of $c^\mu$, we have actually
constructed in \pref{p:Acdress}(i) ``parallel transporters along cones
from $x$ to infinity''.

One may also envisage a generalization with $\pa_\mu c_{x_1,x_2}^\mu(y)
=\delta(y-x_1)-\delta(y-x_2)$ (rather than \eref{pac}),
so that $W(c_{x_1,x_2})$ would qualify as ``parallel transporters
along thickened lines from $x_1$ to $x_2$''. These notions seem not to
exist in mathematical formulations of gauge theory. There is work on
progress by one of us \cite{Hemp} for a coordinate-free formulation on
bundles over manifolds. A somewhat similar (albeit completely
  delocalized) construction arises in \cite{GHR} in terms of the
  Singer-deWitt connection. 

Functions like $c_{x_1,x_2}$ also appear in the discussion of
``gauge bridges'' between (external) charges in QED, \cite{BCRV}.
In this context, abelian operators $\gamma(c_{x_1,x_2})$ model the
electromagnetic field between two charges.

\section{The case of QCD}
  \label{s:QCD}

In the previous section, we have solved the Modified Mediating Equation
\eref{MME} for the Yang-Mills interaction densities $L(c)$ and $K$ pertaining to
sQFT and to gauge theory, respectively, given by \eref{L} and \eref{K}.
The solution was given in terms of the field $W(c)=e^{i\gamma(c)}$
and the mediator field $U^\mu=\erw{F^{\mu\nu}\vert \alpha_\nu}$,
where $\gamma(c)$ and $\alpha_\nu$ were computed recursively as power
series in $g$ in \pref{p:Acdress}.

We now turn to QCD, by adding the minimal interactions $\wt L(c)=
g\Erw{A_\mu(c)\Vert j^\mu}$, $\wt K= g\Erw{A_\mu\Vert j^\mu}$. The quark
fields are assumed to transform in some unitary representation $\pi$ of $\gg$,
hence the current components are $j^\mu_a=\ol\psi\gamma^\mu \tau_a \psi$ where
$\tau_a=\pi(\xi_a)$, and $j^\mu = \sum_a j^\mu_a\xi_a$, see
\sref{s:basics}. In particular, we have to tensor $\WW$ and $\WW_s$ by
the Wick algebra of $N=\dim(\pi)$ free quark fields.

In order to establish the equivalence of S-matrices, we want to
solve the Modified Mediating Equation
\bea{MME-QCD}
e^{\omega_{U+\wt U}}(L(c)+\wt L(c)) - (K+\wt K)
= F(\omega_{U+\wt U})(\pa (U+\wt U)) - D[N]
\eea
with the mediator field $U^\mu+\wt U^\mu$, where $\wt U^\mu$ is of
the form
\bea{wU}
\wt U^\mu= \Erw{\beta\Vert j^\mu} \equiv \ol\psi\gamma^\mu \pi(\beta) \psi
\eea
with a field $\beta\in g\SS^\gg[[g]]$ to be determined. 

\paragraph{Notation.}
We henceforth write $\omega \equiv
\omega_U+\omega_{\wt U}$, where $\omega_U$ acts only on Yang-Mills fields,
while $\omega_{\wt U}$ acts only on quark fields and currents (because
$\beta$ is inert). We shall freely alternate between the two equivalent
writings as on the right-hand side of \eref{wU}. As before, we often
suppress the argument $c$, except in order to distinguish $A_\mu(c)$
(the string-localized potential) from $A_\mu$ (the gauge potential).

First, we note that $\omega_{\wt U}$ acts trivially on all Yang-Mills fields
$X\in\WW_s$, because $\beta\in g\SS^\gg[[g]]$ is inert, \dref{d:sinert},
and $j^\mu$ does not act at all on $\WW_s$. Therefore, $e^{\omega}(L(c))
=e^{\omega_U}(L(c))$ and $F(\omega)(\pa U) = F(\omega_U)(\pa U)$,
and, after subtraction of \eref{MME}, \eref{MME-QCD} reduces to  
\bea{MME-QCDr}
e^{\omega}(\wt L(c)) - \wt K = F(\omega)(\pa \wt U),
\eea
In parallel to \sref{s:main}, we begin by an ansatz \eref{psidress} for the
dressed quark field $e^\omega(\psi)$. The ansatz is motivated by the
gauge transformation $\psi(x)\mapsto \psi^\theta(x)= \pi(\theta(x)\inv)\psi(x)$,
so that  $e^{\omega}(\psi)$ is gauge invariant by \pref{p:GT-W}.
What is more, it allows us to determine the field $\beta$ in \eref{wU}
in terms of $\omega_U$ (hence $\alpha$) and $\gamma$ known from
\sref{s:main}. Then, we show that \eref{MME-QCDr} is satisfied.
We shall need the structural results for the string-localized
obstruction calculus proven in \aref{a:obst-s}.

\begin{prop}\label{p:psidress} {\rm(Dressed quark field and
        reconstruction of the mediator field of QCD.)}
Let $W(c)=e^{i\gamma(c)}$ and $U^\mu=\erw{F^{\mu\nu}\vert \alpha_\nu(c)}$ 
    as in \pref{p:Acdress}. There exists a unique mediator field
    $\wt U^\mu(c)$ of the form \eref{wU} with $\beta(c)\in g\SS^\gg[[g]]$,
    such that $U^\mu+\wt U^\mu$ generates the dressing transformation 
    \bea{psidress}
    e^{\omega}(\psi(x)) &=& \pi\big(W(x,c)\big)\psi(x) \qquad
    (\omega\equiv\omega_U+\omega_{\wt U}). 
    \eea
    In particular, the dressed quark field \eref{psidress} is gauge invariant.
\end{prop}
\textit{Proof.}
We shall determine the field $\beta\in\SS^\gg[[g]]$ in terms of the known
field $\gamma$ and map $\omega_U$.

For $X$ in the polynomial envelopping algebra of $\SS^\gg[[g]]$, we
  have $e^\omega(\pi(X)\psi)=\pi(e^{\omega_U+R_{i\beta}}(X))\psi$ by
  \eref{Rbeta-n}. On the other hand, $e^\omega(\pi(X)\psi)=
  \pi(e^\omega(X))e^{\omega}(\psi)= \pi(e^{\omega_U}(X)e^{i\gamma})\psi$.
  Because $\pi$ is arbitrary, we conclude
  $$e^{\omega_U+R_{i\beta}} = e^{R_{i\gamma}}\circ e^{\omega_U}$$
  as maps on the envelopping algebra.
  
Using the Baker-Campbell-Hausdorff formula on the right-hand side, we
  identify the exponents:
  \bea{BCH}
  \omega_U+R_{i\beta} =
  R_{i\gamma}+ \omega_U + Q(R_{i\gamma},\omega_U),
  \eea
where $Q(X,Y)= \frac12[X,Y]+\frac1{12}([X,[X,Y]]+[Y,[Y,X]]) -
\frac1{24}[X,[Y,[X,Y]]] + \dots$ is a series of $n$-fold
commutators ($n\geq1$). We then use for $s,t\in \SS^\gg[[g]]$
$$[\omega_U,R_{is}]=R_{i\omega_U(s)} \quad \hbox{and}\quad
[R_{is},R_{it}] = -iR_{i[s,t]},$$
to prove by induction in $n$, that every $n$-fold commutator of
$\omega_U$ and $R_{i\beta}$ is a linear combination of $k$-fold
commutators among $R_{i\beta^{(n_i)}}$ ($k\geq0, n_i\geq 0,
k+\sum n_i=n$), where $\beta^{n_i}\equiv\omega_U^{n_i}(\beta)$
is an element of $g^{n_i+1}\SS^\gg[[g]]$ by \lref{l:omSV}.

Thus, $Q(R_{i\gamma},\omega_U)$ is of the form $R_{iq}$,
  where $q\in
  g^2\SS^\gg[[g]]$ can be computed explicitly in terms of the BCH
  expansion. We conclude from \eref{BCH} that
  $R_{i\beta} = R_{i(\gamma +q)}$ as maps on $\SS^\gg$, and
because $R_{is}=R_{it}$ implies $s=t$, we have
\bea{betagamma}
\beta =
  \gamma + q = \gamma -\sfrac12\omega_U(\gamma) +
  \sfrac1{12}\big(i[\gamma,\omega_U(\gamma)] +
  \omega_U^2(\gamma)\big)+\dots.
  \eea
  One can prove by induction that $\beta$ is a series in multiple
  commutators among $\omega_U^{n_i}(\gamma)$. 
  Since $\gamma$ and $U^\mu$ were already obtained in \sref{s:main}
  as power series in $g$ (both of leading order $\OO(g)$), one finally
  gets $\beta$ as a power series in $g\SS^\gg[[g]]$.
  
  Gauge invariance of \eref{psidress} follows from \pref{p:GT-W}. \qed

At lowest orders:
  $$
  \beta = g\gamma_1 +\sfrac{g^2}2\big(\gamma_2-\omega_{U_1}(\gamma_1)\big)
  +\sfrac{g^3}6\big(\gamma_3-\sfrac32\omega_{U_1}(\gamma_2)
  -\sfrac32\omega_{U_2}(\gamma_1)+\sfrac12\omega_{U_1}^2(\gamma_1)
+ \sfrac12i[\gamma_1,\omega_{U_1}(\gamma_1)] 
\big) + \dots ,
$$
where $\omega_{U_n}$ is the contribution from
$U^\mu_n=\erw{F^{\mu\nu}\vert \alpha_{n,\nu}}$, according to
\eref{Ualpha}, which can be worked out with \lref{l:omU}.
\begin{coro} \label{c:jdress}
Along with \eref{psidress}, it also holds
\bea{opsidress}
e^\omega(\ol\psi)(x) &=& \ol\psi(x)\pi(W(x,c)\inv),
\\ \label{jdress}
  e^\omega(j^\mu)(x) &=& W(x,c)j^\mu(x)W(x,c).
  \eea
  \end{coro}
\textit{Proof.} \eref{opsidress} follows from \eref{psidress} because
$\beta$ and $\gamma$ are hermitean fields, so that  $W(c)^*=W\inv$.
\eref{jdress} follows by
$$
e^\omega(j^\mu)=\sum_a\ol\psi\gamma^\mu \pi(W\inv \xi_a W)\psi \xi_a =
  \sum_b\ol\psi\gamma^\mu \pi(\xi_b )\psi W\xi_bW\inv = Wj^\mu
  W\inv. \eqno{\square}
  $$
\begin{coro} \label{c:QCD} {\rm (Equality of S-matrices for QCD.)}
  The equality \eref{SNS} of S-matrices holds for the case of QCD.
\end{coro}
  \textit{Proof.}
We have to establish \eref{MME-QCDr} which would imply \eref{MME-QCD},
so that \lref{l:MME} applies. We shall need the structural results in
\aref{a:QCD}.

By the dressing transformations \eref{Acdress} of $A(c)$ and
\eref{jdress} of $j$, 
\bea{fail}
e^{\omega}(\wt L) = g\Erw{W(A_\mu + i g\inv  \pa_\mu)W\inv\Vert Wj^\mu W\inv}
= \wt K - i\Erw{W\inv\pa_\mu W\Vert j^\mu}.
\eea
Thus, for \eref{MME-QCDr}, it remains to show that
\bea{MME-QCDrr}
F(\omega)(\pa \wt U) \stackrel!= -i \ol\psi\gamma^\mu\pi(W\inv\pa_\mu W)\psi
\equiv -i\Erw{W\inv\pa_\mu W\Vert j^\mu}.
\eea
On the left-hand side, we use that the free current in \eref{wU} is conserved,
hence $\pa \wt U= \erw{\pa_\mu\beta\vert j^\mu}$, and apply \eref{id1},
lifted to arbitrary powers of $\omega$. This yields the equivalent condition 
\bea{cond}
F(\omega_U-\ad_{i\beta})(\pa_\mu\beta) \stackrel!= -i W\inv\pa_\mu W.
\eea
In order to establish \eref{cond}, we exploit that $F(x)=\frac{e^x-1}x =
\int_0^1 dt\, e^{tx}$, and compute 
$$
F(\omega_U-\ad_{i\beta})(\pa_\mu\beta) = \int_0^1 dt\,
e^{t(\omega_U-\ad_{i\beta})}(\pa_\mu\beta) = 
\int_0^1 dt\, W_t\inv \pa_\mu (e^{t\omega_U}(\beta))W_t .
$$
In the second step, we used \eref{id2} and \lref{l:omU}(ii).
By \eref{dtWt} (with $\dot W_t\equiv\pa_tW_t$), this equals
$$
=   \int_0^1 dt\, W_t\inv \pa_\mu \dot W_t  + \int_0^1 dt\,
W_t\inv \dot W_t \pa_\mu W_t\inv W_t =  \int_0^1 dt\,
\pa_t(W_t\inv\pa_\mu W_t) = W\inv \pa_\mu W
$$
(because $W_0=1$). This proves the \eref{cond}, hence \eref{MME-QCDrr},
hence \eref{MME-QCDr}, hence \eref{MME-QCD}.
\qed
\begin{remk}\label{rk:wK0}
In contrast to \eref{K0LK}, adding the free Dirac Lagrangian $\wt K_0=
\ol\psi(\gamma^\mu i\pa_\mu-m)\psi$, which $=0$ in the Wick algebra
of the quark fields by the equation of motion, one does \textit{not} have
$e^{\omega}(\wt K_0+\wt L) = \wt K_0+\wt K$. The reason for the discrepancy
is that the dressing transformation, unlike in \eref{Fdress}, does not
commute with the derivative of the free Dirac field (obvious from
\eref{psidress}: $e^\omega(\gamma^\mu\pa_\mu\psi)=e^\omega(m\psi) =
mW\psi\neq \gamma^\mu\pa_\mu e^\omega(\psi)
=\gamma^\mu\pa_\mu (W\psi)$). It is here that the comment in
\fref{f:quotient} comes to bear: Namely, gauge invariance of the Dirac
Lagrangian $\wt K_0+\wt K$ (as a functional on configurations) is
incompatible with the equation of motion $\wt K_0=0$ which holds only on
the stationary points of this functional. 
\end{remk}
\begin{remk}\label{rk:repn}
The dressing formula \eref{psidress} holds for matter fields $\Phi(x)$ in
arbitrary unitary representations $\pi$, and by the multiplicative property
of the dressing map $e^\omega$ also for composite fields. Thus,
$\gg$-neutral fields (like baryons and mesons, for which $\pi(\gamma(c))=0$,
$\pi(W(c))=1$) are invariant under the dressing. In particular, they
remain local and are insensitive to infrared effects caused by the dressing
factor $W(c)$, cf. \rref{rk:IR}(iii).
\end{remk}

\section{Conclusion}
\label{s:conc}

The S-matrix of string-localized YM and QCD was shown to be
string-independent in previous work \cite{Gass,GGM}. We showed here that it
is actually equivalent to that of YM and QCD treated as gauge theories.
This is achieved by establishing the ``Mediating Equation''
(\cref{c:ME}, \lref{l:MME}) as a sufficient condition, and showing
that it has a solution.

This means that the notorious unphysical artifacts of gauge theory quantization
can in fact be avoided from the outset, rather than being eliminated
\textit{ex post}. A reason why this is so has been identified along the way: it
is the gauge invariance of the underlying \textit{classical} Lagrangian. That
gauge transformations need not be implemented in the \textit{quantum}
field theory, is the main conceptual result of the paper.

The Mediating Equation is solved by an ``educated guess''
  concerning the form of the dressed fields $e^{\omega_U}(\Phi)$
  of the sQFT approach. The self-consistency of this solution follows
  because it allows to reconstruct the mediator field $U^\mu$ featuring
  in the Mediating Equation.

The dressed fields, for which exact formulas are given in this paper,
are expected to anticipate salient infrared features of the
interacting fields. This is known in the case of QED, 
and we speculate for QCD that their correlations may already encode
``kinematic'' indications of confinement, see \rref{rk:IR}(iii). Validating this
expectation appears as a most promising objective for further research.

The present results are at tree-level only, but their preservation at loop
level may be considered as renormalization conditions for loops with
string-localized propagators. This issue is still open. The ``enhanced
obstruction calculus'' introduced and applied in this work for the
purpose of establishing equivalence, will presumably not be needed for
the actual perturbation theory at loop level. It is rather expected to
implicitly care for cancellations among string-dependent contributions to loops.

\appendix

\section{Obstruction calculus for string-integrated
  Wick products}
\label{a:obst-s}

\subsection{Yang-Mills sector}
\label{a:YM}

We have to extend \eref{OYXexp} to $Y$ and $X\in\WW_s$, defined in \sref{s:Wick}. As for
$X\in\WW_s$, we observe that all entries in \tbref{tb:YM} are
compatible with the rule 
\bea{OI}
O_\mu(Y',I^\ka_cX)=I_c^\ka O_\mu(Y',X).
\eea
We may extend it to hold also for Wick polynomials $X$, because the
derivatives involved in the operation $O_\mu(Y',\dots)$ do not
interfere with the string-integration over $x$. This settles the
extension to $X\in\WW_s$.

As for $Y\in\WW_s$, when $Y=I(Y_1)Y_2$, one may use \eref{OYXexp}
recursively, provided $O_\mu(I'(Y_1'),X)$ is known. However, when
$Y=I(Y_1Y_2)$, trouble is ahead. Trying to define also
$O_\mu(I'^\nu_cY',X):=I'^\nu_c O_\mu(Y',X)$ so that \eref{OYXexp} can
be used, leads to inconsistencies.
E.g., there is no way to define $\vev{T\pa'_\mu A'_\nu(x',c)\ndot F^{\ka\la(x)}}$
(needed to compute $O_\mu(A'_\nu(c),F^{\ka\la})$), that would be
compatible with the rows 2 and 3 of \tbref{tb:YM}, {\em and} with
$I'^\nu_cO_\mu(A'_\nu(c),F^{\ka\la})\stackrel!=O_\mu(I'^\nu_cA'_\nu(c),F^{\ka\la})=0$.
\footnote{This can be easily seen because it must be symmetric in
  $\mu,\nu$ because of row 2 of \tbref{tb:YM}, and therefore must be a
  linear combination of the three structures 
  $f_1\eta_{\mu\nu}I'^{[\ka}\pa^{\la]}i\delta(x'-x)
  +f_2I'_{(\mu}\delta_{\nu)}^{[\ka}\pa^{\la]}i\delta(x'-x)
  +f_3\pa_{(\mu}\delta_{\nu)}^{[\ka}I'^{\la]}i\delta(x'-x)$. The
vanishing of $O_\mu(A'_\nu(c),F^{\ka\la})$ would require
$f_1=f_2=f_3=0$, which is inconsistent with row 4 of \tbref{tb:YM}.}
The result is perhaps not too surprising because, in view of
\eref{Ipa}, time-ordering not commuting with the derivative $\pa'$,
should also not unreservedly commute with $I'_c$.

As a consequence, for fields of the form $s= I_c(i[Y_1,Y_2])$,
the operation $O_\mu(\sigma',\cdot)$ is not a priori defined by \eref{OYXexp}.
However, when $s\in\SS^\gg$, the convention
$T\pa'_\mu \sigma(x')\ndot X(x):=\pa'_\mu T \sigma(x')\ndot X(x)$ as a
choice of renormalization of time-ordered products is justified
because $s\in\SS^\gg$ do not satisfy equations of
motion. The
only exception is $\phi\in\SS^\gg$ for which time-ordered products of $\pa\phi =
A(c)-A$ are independently given. But $\phi$ is inert, nonetheless, by \tbref{tb:YM}.

To wrap this up, we are free to postulate inertness of $\SS^\gg$ as a
choice of renormalization:
\begin{defn}\label{d:sinert}
  We define
  \bea{sinert}
  O_\mu(s(x'),X(x)):=0 \quad\hbox{for all $s\in \SS^\gg$, $X\in\WW_s$}.
  \eea
\end{defn}
We now prove several consequences of this choice, that are crucial in \sref{s:main}.

\begin{lemma}\label{l:Owv}
  Let $v\in \VV^\gg$, $X\in\WW_s$. Then
  $$
  O_{[\mu}(v_{\nu]}(x'),X(x))=0,
  $$
 i.e., $v$ is ``inert'' in the operation
 $O_{[\mu}(v'_{\nu]},\cdot)$. The same is true for $v\in\VV^\gg[[g]]$.
\end{lemma}
{\em Proof:}
Consider the two types of elements of $\VV^\gg$ in
\dref{d:SV}(ii). Because all $s_k$ are inert by \dref{d:sinert},
  it suffices to show that
$O_{[\mu}(A'_{\nu]},X)=0$ and
$O_{[\mu}(\pa'_{\nu]}s_0',X)=0$. The former is true
by row 4 of \tbref{tb:YM}. For the latter, we use that $T\pa'_\nu s(x')\ndot
X(x)=\pa'_\nu Ts(x')\ndot X(x)$ by \dref{d:sinert}:
$$
O_{[\mu}(\pa'_{\nu]}s(x'),X(x)) = T\pa'_{[\mu}\pa'_{\nu]}s(x')\ndot
X(x)-\pa'_{[\mu}T\pa'_{\nu]}s(x')\ndot
X(x)=-\pa'_{[\mu}\pa'_{\nu]}Ts(x')\ndot X(x) =0 . 
$$
The extension to power series is obvious by linearity of
$O_\mu(U'^\mu,X)$ in the first entry. \qed
\begin{lemma}\label{l:omU}
  Let $\alpha\in\VV^\gg$, $U^\mu:=\erw{F^{\mu\nu}\vert \alpha_\nu}$, and $X\in \WW_s$.
  \\
  (i)
  It holds
  \bea{omegaUX}
  \omega_U(X)(x)= \int dx'\, \Erw{\alpha_\nu(x')\Vert
    O_\mu(F^{\mu\nu}(x'),X(x))}.
  \eea
In particular,
  $\omega_U(N)= \omega_U(\pa^\ka N)=0$ and
  \bea{omFAphi}
  \omega_U(F_{\mu\nu})=\pa_{[\mu}\alpha_{\nu]} \quad
  \omega_U(\phi)=(I_c^\nu\alpha_\nu),\quad
  \omega_U(A_\mu)=\alpha_\mu, \quad
  \omega_U(A_\mu(c))=\alpha_\mu+\pa_\mu(I_c^\nu\alpha_\nu). \quad
  \eea
In particular, replacing $\alpha_\mu$ by $\alpha_\mu + \pa_\mu s$ does
not affect $\omega_U(F)$ and $\omega_U(A(c))$.
\\[1mm] 
(ii)
For $s\in \SS^\gg$ and $v\in \VV^\gg$, it holds
$$
\omega_U(\pa_\mu s)=\pa_\mu\omega_U(s), \qquad
\omega_U(\pa_{[\mu}v_{\nu]})= \pa_{[\mu}\omega_U(v_{\nu]}).
$$
(iii)
For $v\in \VV^\gg$, it holds 
$$
\omega_U(I_c^\nu v_\nu )=I_c^\nu\omega_U(v_\nu),
$$
and in combination with (ii), also $\omega_U((\delta_\mu^\nu +
  \pa_\mu I_c^\nu)(v_\nu))=(\delta_\mu^\nu + \pa_\mu
  I_c^\nu)\omega_U(v_\nu)$.
  \\
(i)--(iii) are true as well for $\alpha\in \VV^\gg[[g]]$, $s\in\SS^\gg[[g]]$, $v\in\VV^\gg[[g]]$. 
\end{lemma}
{\em Proof:} 
(i) By \lref{l:Owv}, $\erw{F'^{\mu\nu}\vert O_\mu(\alpha'_\mu,X)}=0$.
This implies \eref{omegaUX}.
The evaluation of $\omega_U$ on the linear fields follows from row 1
of \tbref{tb:YM}. At this point, \dref{d:outside} has
to be taken into account: When \eref{omegaUX} appears inside a
T-product, the integration by parts leading to \eref{omFAphi} produces 
errors for $X=F^{\ka\la}$ and $X=A^\ka(c)$, as displayed in \fref{fn:MWI}. The
error is zero in both cases: in the former case 
thanks to \lref{l:omU}(i) (since $\alpha\in \VV^\gg$), and in the latter case it is
$$\sim \int dx'\, O^\ka (I^\nu\delta(x-x')\alpha_\nu(x'),\cdot)
=O^\ka (I_c^\nu\alpha_\nu(x),\cdot)=0$$
by \dref{d:sinert} (since $I^\nu_c\alpha_\nu\in \SS^\gg$).
\\[1mm]
The first equation of (ii) follows from $[[T,\pa'_\mu],\pa_\ka]=[[T,\pa_\ka],\pa'_\mu]$
(in the short-hand notation as in \eref{OYX}), implying
$$
O_\mu(U'^\mu,\pa_\ka s)- \pa_\ka O_\mu(U'^\mu,s) = O_\mu(s,\pa'_\mu
U'^\mu) - \pa'_\mu O_\mu(s,U'^\mu).
$$
This vanishes because $s$ is inert, \dref{d:sinert}. The second
equation follows in the same way, using \lref{l:Owv} in the 
last step.
\\
For (iii), recall that \eref{OI} extends to all $X\in \WW_s$, as
discussed above.
\\
The extension to power series is again obvious by linearity of
$O_\mu(U'^\mu,X)$ in the first entry. 
\qed
\begin{lemma}\label{l:omSV}
  For $\alpha\in\SS^\gg$, $U^\mu=\erw{F^{\mu\nu}\vert \alpha_\nu}$, it holds
  $$
  \omega_U(\SS^\gg)\subset \SS^\gg, \qquad \omega_U(\VV^\gg)\subset \VV^\gg.
  $$
\end{lemma}
\textit{Proof.}
By induction in $k$ in \dref{d:SV}, using the derivation property of
$\omega_U$ along with \tbref{tb:YM} and \lref{l:omU}(i)--(iii).
\qed 
  \begin{remk}\label{rk:MWI} The proof of \lref{l:omU}(i) shows
    that the map $\omega_U$, where $U^\mu$ is of the form \eref{Ualpha},
    is insensitive to the subtlety of the MWI addressed in \dref{d:outside}
    (and which necessitated the Erratum to \cite{LV}). For
    QCD, the same is true because derivatives of $\delta$-functions do
    not appear in $O_\mu(\wt U'^\mu,X)$.
  \end{remk}

\subsection{Quark sector}
\label{a:QCD}

We first compute
$$
i\vev{T\gamma^\mu\pa'_\mu\psi'\ndot\ol\psi} -
i\gamma^\mu\pa'_\mu \vev{T\psi'\ndot\ol\psi} =
(m-i\gamma^\mu\pa'_\mu)\vev{T\psi'\ndot\ol\psi} = -i\delta(x-x'),
$$
implying
\bea{no4}
\notag iO_\mu(j^\mu_a(x'),\ol\psi(x)) &=&
-i\ol\psi(x')\pi(\tau_a)\delta(x'-x),
\\ \notag
\hbox{and likewise} \quad  iO_\mu(j^\mu_a(x'),\psi(x)) &=&
i\pi(\tau_a)\psi(x')\delta(x'-x).
\eea
Consequently, for $\wt U$ of the form given in \eref{wU}
\bea{omopsi} 
 \omega(\ol\psi)(x) & =& \omega_{\wt U}(\ol\psi)(x) =
-i\ol\psi(x)\pi(\beta(x,c)),
\\ \label{ompsi}
  \omega(\psi)(x) & =&
 i\pi(\beta(x,c))\psi(x).
 \eea
For $X$ in the polynomial envelopping algebra of
$\SS^\gg[[g]]$, \eref{ompsi} implies
\bea{Rbeta}
\omega(\pi(X)\psi) = \pi(\omega(X))\psi + i\pi(X\beta)\psi =
\pi((\omega_U+R_{i\beta})(X))\psi,
\eea
where $R_{i\beta}$ is the right multiplication with $i\beta$. By iteration
\bea{Rbeta-n}
\omega^n(\pi(X)\psi) = \pi((\omega_U+  R_{i\beta})^n(X))\psi.
\eea
In particular, $(\omega_U+R_{i\beta})^n(1)$ is a polynomial of elements
$\omega_U^{n_i}(\beta)\in g^{n_i+1}\SS^\gg[[g]]$ ($n_i<n$). Summing these
polynomials, we define
  \bea{Wt}
  W_t(x,c) := e^{t(\omega_U+R_{i\beta(x,c)})}(1).
  \eea
  \begin{lemma}\label{l:coc} {\rm (Dressing cocycle.)}
(i) It holds
\bea{Wtpsi}
e^{t\omega}(\psi)(x) = \pi(W_t(x,c))\psi(x).
\eea
(ii) The map $\RR\ni t\mapsto W_t(c)$ satisfies the cocycle relation
$W_{t_1+t_2} =  e^{t_2\omega}(W_{t_1})W_{t_2}$ and  
  \bea{dtWt}
 \pa_t W_t(x) = e^{t\omega_U}(i\beta)(x) W_t(x).
\eea
In particular, $W_t \in\exp ig\SS^\gg[[g]]$, $W_0=1$, and $W_1(x,c)=W(x,c)$ in \eref{Wgamma}.
\end{lemma}
\textit{Proof.} \eref{Wtpsi} is an immediate consequence of \eref{Rbeta-n}. The
cocycle relation and \eref{dtWt} follow by application of
$e^{t_2\omega}$ and $\pa_t$, respectively, to
\eref{Wtpsi}. Defining $W_t=:e^{i\gamma_t}$, one can rewrite \eref{dtWt}
as $F(\ad_{i\gamma_t})(\dot\gamma_t) = e^{t\omega_U}(\beta)$, which implies
$\gamma_t\in g\SS^\gg[[g]]$. \eref{psidress} and \eref{Wtpsi} imply $W_1=W$.
\qed
\begin{lemma} \label{l:id12}
  For $X\in \WW^\gg_s$, it holds
  \bea{id1}
  &&\omega\big(\Erw{X\Vert j^\mu}\big) =
  \Erw{(\omega_U-\ad_{i\beta})(X))\vert j^\mu},
  \\ \label{id2}
  && W_t\inv e^{t\omega_U}(X) W_t =
  e^{t(\omega_U-\ad_{i\beta})}(X). 
  \eea
\end{lemma}
\textit{Proof.} \eref{id1} is proven like \eref{Rbeta}, writing
$\erw{X\vert  j^\mu}=\ol\psi\gamma^\mu\pi(X)\psi$ and taking
\eref{omopsi} and \eref{ompsi} together. Lifting \eref{id1}
to arbitrary power series in $\omega$, we also have 
$$
e^{t\omega}\big(\Erw{X\Vert j^\mu}\big) =
\Erw{e^{t(\omega_U-\ad_{i\beta})}(X))\vert j^\mu}.
$$
On the other hand, the multiplicative property of $\omega$ implies
$$
e^{t\omega}\big(\Erw{X\Vert j^\mu}\big) = e^{t\omega}(\ol\psi\gamma^\mu \pi(X)
\psi) = \ol\psi\gamma^\mu \pi(W_t\inv e^{t\omega_U}(X) W_t) \psi
= \Erw{W_t\inv e^{t\omega_U}(X) W_t\Vert j^\mu}.
$$
This yields \eref{id2}.
\qed

\section{Comparison to the ``Dressing Field Method''}
\label{a:DFM}
Let us point out a surprisingly close
\footnote{We became aware of the ideas of DFM only recently, when the
  general picture of our results was already developped.} 
 relationship between the ``Dressing Field
Method'' (DFM) for classical gauge theories, see \cite{DFM}, as well as some
pertinent differences.

DFM strives to reformulate classical gauge field theories in terms of
  gauge-invariant ``dressed fields''. (The reader should be warned
  that the meaning of this term is not identical in DFM \cite{DFM} and
  in sQFT \cite{AHM, LV}. E.g., the sQFT dressing of $A(c)$ coincides
  with the DFM dressing of $A$, see \cref{c:Wuinv} below.) We
  outline the idea in terms of the classical gauge potential
\footnote{The connection ``$A$'' in \cite{DFM} is in fact $-ig
  A$ in the present paper.}
 $A$ with the gauge transformation 
  \bea{GT}
  A(x)\mapsto A^\theta:=\theta(x)\inv(A(x)+ ig\inv \pa)\theta(x), \qquad \theta(x)\in G.
  \eea
DFM \textrm{assumes} the existence of a group-valued ``dressing field''
$u$ which is a function of $A$, transforming under \eref{GT} as
\bea{GT-u}
u(x) \equiv u[A(x)]\mapsto u[A^\theta(x)]\stackrel!=\theta(x)\inv u(x).
\eea
Then it follows that
\bea{Au}
A^u:=u[A]\inv(A(x)+ig\inv \pa) u[A]
\eea
is gauge invariant.
\footnote{It is emphasized in \cite{BF} that the dressing
transformation \eref{Au}, inspite of its formal congruence with
\eref{GT} is \textit{not} a gauge transformation, because $u$ is not a
group-valued function that acts on every point in a fibre in the
same way, but a group-valued functional of $A$ that brings every
point in the fibre to a gauge fixed section.}

Identifying $W(c)=W[A,I_c]\in\exp ig\SS^\gg[[g]]$ and the free quantum
 quark field $\psi$ with the corresponding classical fields, we
conclude from \pref{p:GT-W}:
\begin{coro}\label{c:Wuinv}
Under the said identification, the group-valued field $W(c)\inv=W[A,I_c]\inv$
qualifies as a dressing field $u$ in the sense of the Dressing Field Method.
With $u=W(c)\inv$, the dressed fields in the sense of sQFT
$e^{\omega_U}(A(c))=A^{W(c)\inv}$ in \eref{Acdress} and
$e^{\omega}(\psi) = \pi(W(c))\psi$ in \eref{psidress} are invariant
under classical gauge transformations, and coincide with the dressed
fields $A^u$ and $\psi^u$ of DFM. 
\end{coro}
In contrast to DFM, where the existence of a classical dressing field $u[A]$
as in \eref{GT-u} is \textit{assumed}, sQFT \textit{constructs} infinitely
many such fields (for any $I_c$ with arbitrary $c$ as in \eref{Ic}). While
the resulting gauge-invariant dressed fields are the same in both
approaches, the dressing \textit{maps} differ in the YM case where
the respective undressed fields ($A(c)$ versus $A$) are not the same.

\paragraph{Acknowledgment.} We thank Michael Dütsch for
  pointing out the subtleties of the on-shell Master Ward Identity.
KHR thanks Ma\"el Chantreau for assistance at a very early stage of
this work. We are indebted to the peer reviewer for very
  constructive comments that helped to improve the manuscript.



\end{document}